\documentclass[11.5pt, review]{elsarticle}

\usepackage{etoolbox}
\makeatletter
\patchcmd{\ps@pprintTitle}
{Preprint submitted to}
{Accepted version in}
{}{}
\makeatother

\usepackage{lineno,hyperref}
\modulolinenumbers[5]

\journal{Aerospace Science and Technology}

\usepackage{geometry}
\usepackage{graphicx}
\usepackage{amsmath}
\usepackage{amssymb}
\usepackage{lineno}

\usepackage{epstopdf}
\usepackage[linesnumbered, ruled]{algorithm2e}
\usepackage{nomencl}
\makenomenclature

\makeindex

\renewcommand\appendix{\par
	\setcounter{section}{0}%
	\setcounter{subsection}{0}%
	\gdef\thetable{\arabic{table}}%
	\gdef\thesection{\appendixname\Alph{section}}%
}

\begin{document}
	
\begin{frontmatter}
	\title{Chance-constrained Model Predictive Control for Near Rectilinear Halo Orbit Spacecraft Rendezvous}
	\author[1]{Julio C. Sanchez} 
	\author[1]{Francisco Gavilan}
	\author[1]{Rafael Vazquez\corref{cor1}}
	\ead{rvazquez1@us.es}
	\address[1]{Departamento de Ingenier\'ia Aeroespacial, Escuela T\'ecnica Superior de Ingenier\'ia, Universidad de Sevilla, 41092, Sevilla, Spain}
	
	\cortext[cor1]{Corresponding author}
	
	
	\begin{abstract}
		This work presents a robust Model Predictive Controller (MPC) to solve the problem of spacecraft rendezvous in the context of the restricted three-body problem (R3BP) as will be required to dock with space stations in cislunar space. The employed methodology is both valid for chemical and electric thrusters. By exploiting the state transition matrix and using a chance-constrained approach, the robust MPC assures constraints satisfaction under the presence of disturbances in a probabilistic sense. The perturbations parameters are computed on-line using a disturbance estimator. The robust controller is tested for a rendezvous scenario with a target placed in an Earth-Moon $L_2$ Near-Rectilinear Halo Orbit. Numerical results are shown and discussed.
\end{abstract}
\end{frontmatter}

\section{Introduction}

Demonstrating rendezvous capabilities in the context of multi-body environments is becoming a growing and active field of research as International Space Station (ISS) partners have interest in building a space station in the cislunar space, named as the Lunar Orbital Platform Gateway (LOP-G), see \cite{Merri2018}. Moreover, this lunar space station will greatly enhance scientific opportunities by allowing to return samples from the Moon, see \cite{Landgraf2018}.

Several options have been studied to place the LOP-G, see \cite{Whitley2016}, being the Near Rectilinear Halo Orbits (NRHOs), around the $L_2$ Earth-Moon point, the most attractive candidates. NRHOs are members of the broader set of $L_1$ and $L_2$ families of Halo orbits existing in the circular restricted three-body problem (CR3BP), see \cite{Doedel2003} for more details about CR3BP orbits. The NRHOs also persist in higher-fidelity models since they present favourable stability properties, see \cite{Zimovan2017}.

Typically, far-rendezvous operations, where fuel consumption is the key driver instead of safety considerations, have been extensively studied in the literature. Reference \cite{Gomez2004} exploits the method of invariant manifolds connections whereas surrogate models, to ease the computational burden of global optimization, have been proposed by \cite{Peng2013}. Finally, \cite{Sato2015} compared the fuel efficiency of classical phasing strategies with invariant manifolds connections.
 
On the other hand, close rendezvous operations (where safety is a main concern) are starting to gain more momentum. Reference \cite{Jones1994} proposed a targeting law combined with a navigation filter for restricted three body problem (R3BP) rendezvous operations. Practical rendezvous scenarios for Earth-Moon Halo orbits were proposed in \cite{Murakami2015}, whereas shooting methods to achieve rendezvous have been studied in \cite{Lizydestrez2019}. The previous works have expressed the system dynamics in the Earth-Moon co-rotating reference frame. However, this frame is not very useful to describe state constraints attached to the target. This is the reason why local frames are widely preferred in close rendezvous operations, see \cite{Fehse2003}. In \cite{Franzini2017}, a local frame of reference is proposed taking into account that the LOP-G will be orbiting the Moon in a practical sense. 

The purpose of this work is to develop a robust rendezvous controller for R3BP scenarios. The key idea behind robust control is to explicitly take into account disturbances and uncertainties in the optimization problem. In the case of Keplerian rendezvous operations, several robust techniques have been explored. Reference \cite{Gavilan2012} employed the chance-constrained approach to guarantee constraints satisfaction probabilistically. A worst-case scenario methodology, to minimize the size of the terminal arrival set, was proposed by \cite{Louembet2015}. Finally, a tube-based method,  guaranteeing constraint satisfaction for bounded disturbances, has been experimentally validated in \cite{Mammarella2018}. 

The main contribution of this work is the extension of the chance-constrained approach, developed in \cite{Gavilan2012}, to R3BP rendezvous. The proposed method explicitly considers the disturbances, affecting the state constraints, in a probabilistic sense. Then, these probabilistic constraints are bounded at a certain probability, which allows to  compute control signals in a deterministic way. Since a priori knowledge of the disturbances statistical properties is required, an on-line estimator of stochastic parameters is also employed. The robust program is embedded into a Model Predictive Control (MPC) scheme, see \cite{Camacho2004}, so the robust program is updated after each sampling time.

Moreover, for this type of mission, the propulsive plant of the chaser can be either chemical or electrical. To extend the potential application of this work, both the impulsive and continuous thrust models are considered. For the continuous thrust case, it is assumed that the control signal can be linearly parameterized by some decision variables. As an additional contribution, basis splines (B-splines), typically employed for attitude control as in \cite{Louembet2009} and \cite{Sanchez2020}, are chosen to parameterize the control signal.  

The structure of this work is as follows. Section 2 describes motion in the restricted three body problem and the linearized relative model. Section 3 follows describing the rendezvous problem. Section 4 formulates the chance-constrained based MPC and the on-line disturbance estimator. Section 5 shows numerical results through a Monte Carlo comparison of the robust and non-robust controllers. Section 6 closes the paper with some final remarks.       

\section{Relative motion in the restricted three body problem}

This section studies the relative motion between two vehicles in the R3BP. Firstly, the motion of a particle, under R3BP assumptions, is described. Additionally, some facts about NRHOs are given. Then, the local-vertical local-horizontal (LVLH) frame is introduced and the R3BP relative dynamics deduced. Finally, the relative motion is linearized assuming that the vehicles are close enough.  

\subsection{Restricted three body problem and NRHOs}

\begin{figure}[h] 
	\begin{center}
		\includegraphics[width=8.5cm,height=8.5cm,keepaspectratio]{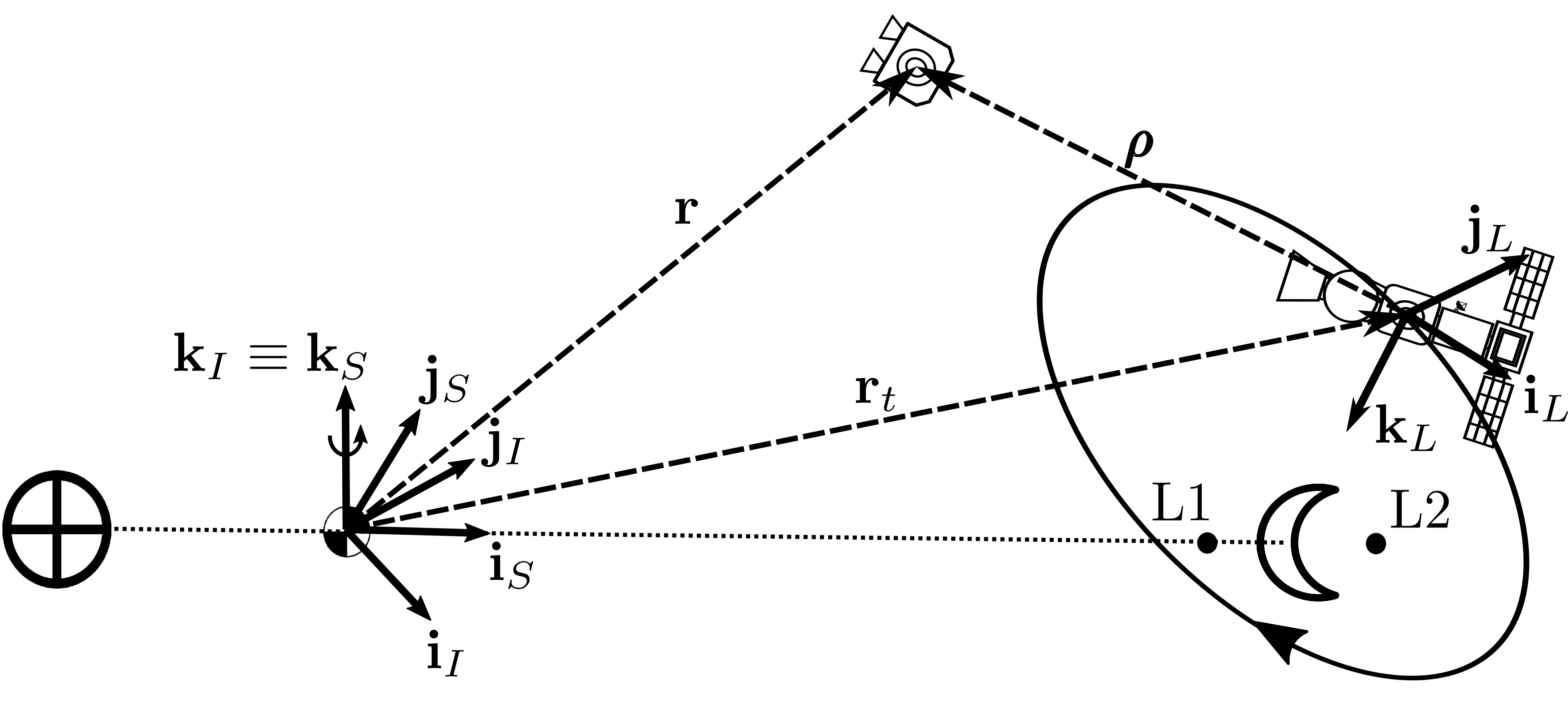}
	\end{center}
	\caption{Inertial, synodic and LVLH frames of reference for the Earth-Moon system.}	
	\label{fig:LVLH_frame}
\end{figure}

Under R3BP assumptions, where $\mu_1 \geq \mu_2 \gg \mu$, being $\mu_1$ and $\mu_2$ the gravitational parameters of the two primaries and $\mu$ that of the vehicle, the spacecraft dynamics are conveniently expressed in the synodic frame, see \cite{Gomez2001}. Denote the inertial frame by $I:\{\mathbf{O},\mathbf{i}_I,\mathbf{j}_I,\mathbf{k}_I\}$ where $\mathbf{O}$ is the position of the system barycenter. Denote the synodic frame by $S:\{\mathbf{O},\mathbf{i}_S,\mathbf{j}_S,\mathbf{k}_S\}$, with $\mathbf{i}_S$ coincident with the line uniting the two primaries and positive in the direction of the second primary, $\mathbf{k}_S$ parallel to the system kinetic momentum and $\mathbf{j}_S$ completing a right-handed system, see Fig.\ref{fig:LVLH_frame}.
The R3BP equations in the $S$ frame are 
\begin{equation}
\begin{aligned}
\left.\mathbf{\ddot{r}}\right|_S=&-\frac{\mu_1(\mathbf{r}-\mathbf{r}_1)}{\|\mathbf{r}-\mathbf{r}_1\|^3_2}-\frac{\mu_2 (\mathbf{r}-\mathbf{r}_2)}{\|\mathbf{r}-\mathbf{r}_2\|^3_2}-2\pmb{\omega}_{S/I}\times\left.\mathbf{\dot{r}}\right|_S
-\left.\pmb{\dot{\omega}}_{S/I}\right|_S\times\mathbf{r} -\pmb{\omega}_{S/I}\times(\pmb{\omega}_{S/I} \times\mathbf{r})+\mathbf{u},
\end{aligned}
\label{eq:R3BP_dyn_synodic}
\end{equation}
where $\mathbf{r}$ is the spacecraft position, $\mathbf{r}_1$ and $\mathbf{r}_2$ the primaries position, $\pmb{\omega}_{S/I}$ the angular velocity of the synodic frame with respect to the inertial  and $\mathbf{u}$ the control acceleration.      

\begin{figure}[h] 
	\begin{center}
		\includegraphics[width=9.5cm,height=9.5cm,keepaspectratio]{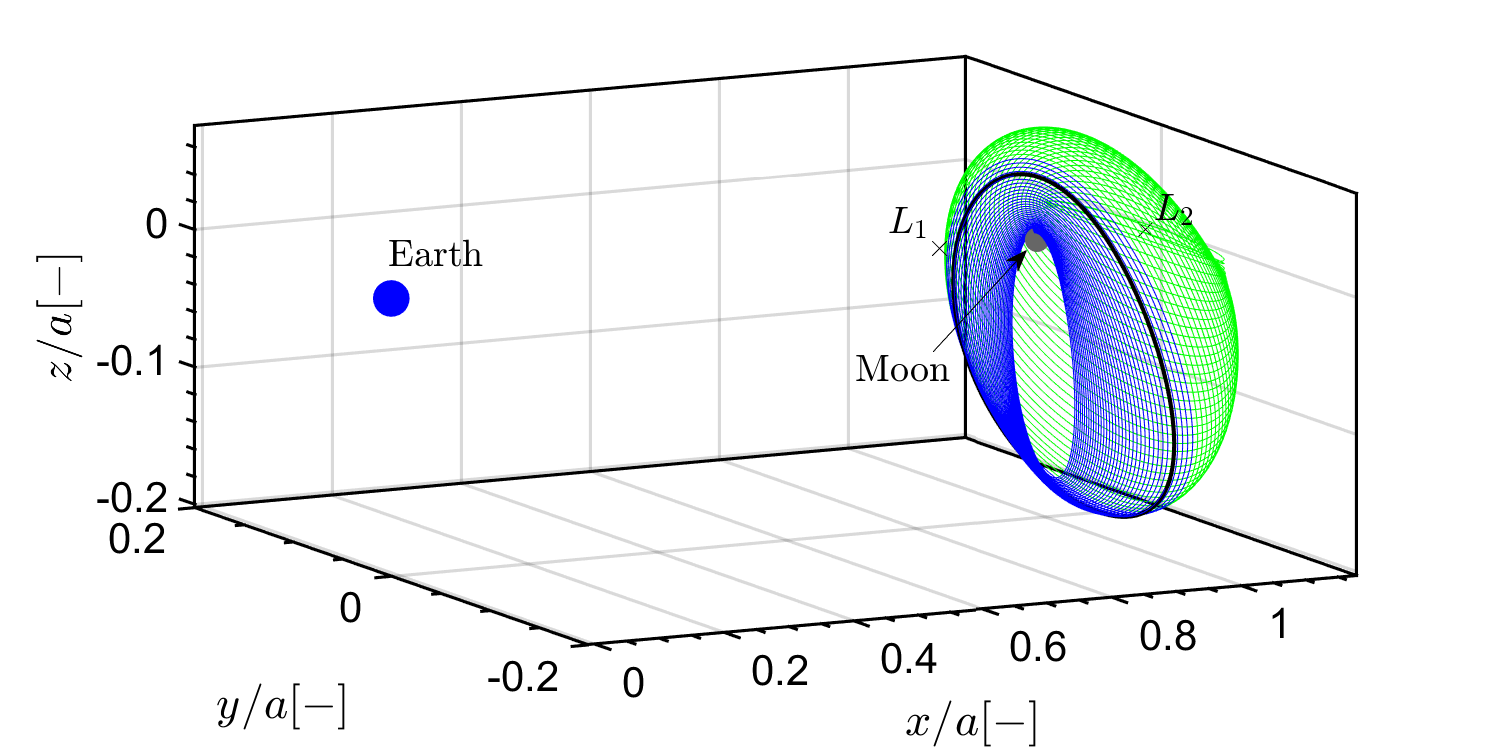}
	\end{center}
	\caption{Green: Southern $L_2$ Halo family; blue: Southern $L_2$ NRHOs; black: Sec.V NRHO. Parameter $a$ is the Earth-Moon semimajor axis.}	
	\label{fig:NRHO_family}
\end{figure}

Eq.\eqref{eq:R3BP_dyn_synodic} allows primaries in elliptic orbits. To obtain the CR3BP equations (circular orbits), set $\pmb{\omega}_{S/I}=n\mathbf{k}_S$ and $\pmb{\dot{\omega}}_{S/I}=\mathbf{0}$ in Eq.\eqref{eq:R3BP_dyn_synodic}, obtaining
\begin{equation}
\begin{aligned}
\left.\mathbf{\ddot{r}}\right|_S=&-\frac{\mu_1(\mathbf{r}-\mathbf{r}_1)}{\|\mathbf{r}-\mathbf{r}_1\|^3_2}-\frac{\mu_2 (\mathbf{r}-\mathbf{r}_2)}{\|\mathbf{r}-\mathbf{r}_2\|^3_2}-2n\mathbf{k}_S\times\left.\mathbf{\dot{r}}\right|_S
-n\mathbf{k}_S\times(n\mathbf{k}_S \times\mathbf{r})+\mathbf{u},
\end{aligned}
\label{eq:CR3BP_dyn_synodic}
\end{equation}
where $n=\sqrt{(\mu_1+\mu_2)/D^3}$ and $D$ is the distance between the two primaries. The CR3BP system \eqref{eq:CR3BP_dyn_synodic} has five libration points, named as Lagrange points ($L_i, \> i=1\hdots5$), with associated families of periodic orbits around them, see \cite{Doedel2003}. Amongst these periodic orbits, the ones receiving more attention, for practical purposes, are the Halo orbits around  collinear equilibria. Since these are unstable, the Halo orbits are in turn inherently unstable, requiring station-keeping to be maintained. Amongst each set of $L_1$ and $L_2$ Halo orbits, there exists a subset (NRHOs) with favourable stability properties. These properties have shown to persist in higher-fidelity models, and hence these orbits may support long-term missions near the Moon. Regarding scientific opportunities, the preferred Earth-Moon NRHOs are the ones associated to the Southern $L_2$ family. This family allows great coverage for both the lunar South pole and far side of the Moon, see \cite{Hartley2015}. Covering these areas is of great scientific interest due to the existence of water ice in the South pole, see \cite{Hayne2015}, and the impossibility to observe the far side of the Moon from Earth. The Southern $L_2$ Halo family and their subset of NRHOs for the Earth-Moon system are shown in Fig.\ref{fig:NRHO_family} in a non-dimensional synodic frame. Note that they can be practically seen as lunar orbits, with the perilune at the North pole.

\begin{figure}[h] 
	\begin{center}
		\includegraphics[width=8.5cm,height=8.5cm,keepaspectratio]{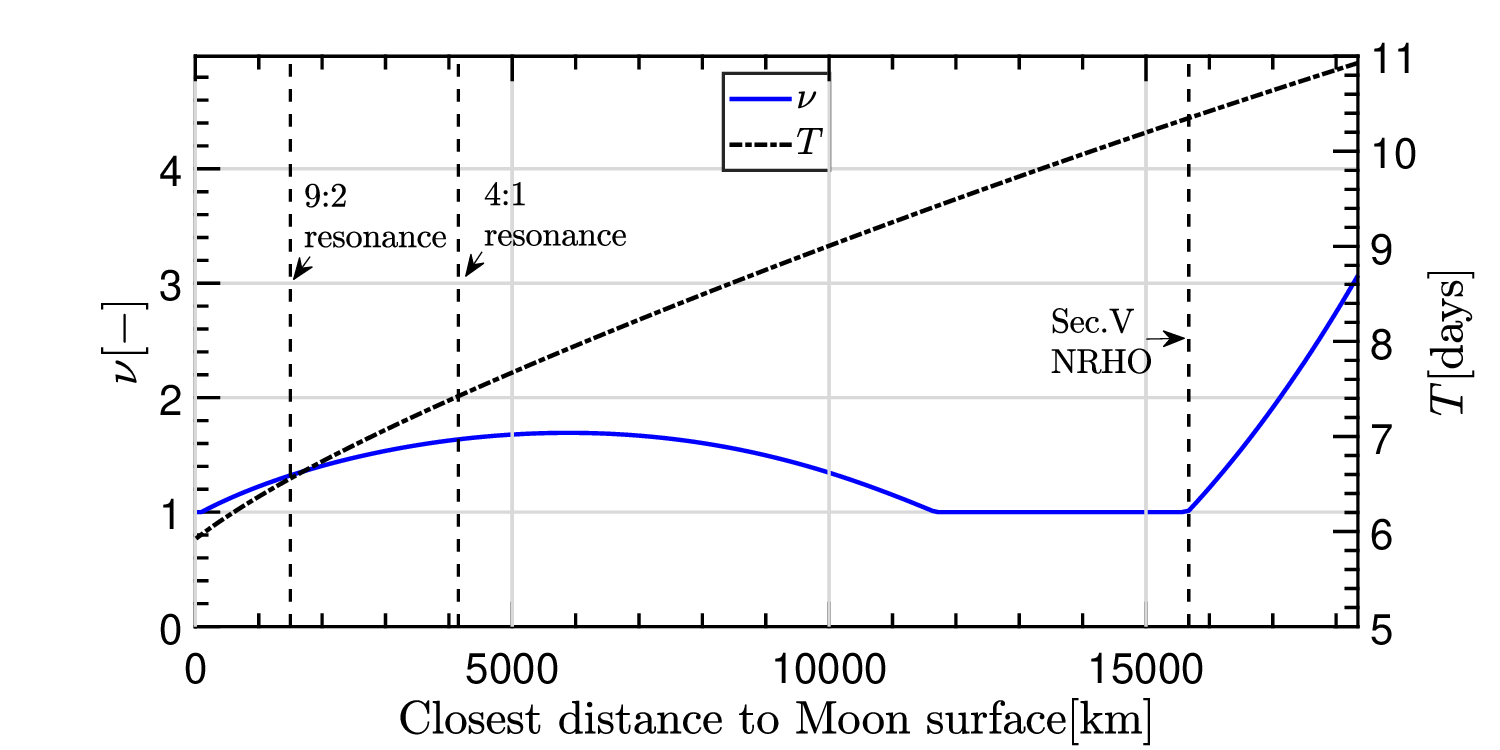}
	\end{center}
	\caption{Stability indexes and periods for Southern $L_2$ NRHOs}	
	\label{fig:NRHO_stability_periods}
\end{figure}

To evaluate the stability properties of CR3BP periodic orbits, \cite{Davis2017} proposed the stability index parameter $\nu$
\begin{equation}
\nu=\frac{1}{2}\left(\lambda_{max}+\frac{1}{\lambda_{max}}\right),
\end{equation}
which is a function of $\lambda_{max}$, the absolute value of the monodromy matrix (state transition matrix after one orbital period) maximum real eigenvalue (in absolute value). The monodromy matrix of an autonomous Hamiltonian system is symplectic, hence each eigenvalue $\lambda$ has an opposite one $\lambda^{-1}$, see \cite{Koon2006} for the details. Since the orbit is periodic, two monodromy matrix eigenvalues are always equal to the unity. As a consequence $\nu\geq1$ and the periodic orbit is marginally stable if $\nu=1$ and unstable if $\nu>1$. Both the stability indexes and orbital periods for the Southern $L_2$ NRHOs are shown in Fig.\ref{fig:NRHO_stability_periods}. It can be seen that at very close distances from the Moon surface, the NRHOs are almost marginally stable. As distance from the Moon increases, the stability indexes rise and decrease until they become almost marginally stable for altitudes ranging from $11500~\text{km}$ to $16750~\text{km}$. Afterwards the stability indexes begin to increase quickly becoming highly unstable. Additionally, Fig.\ref{fig:NRHO_stability_periods} shows the period increases monotonically with respect to the perilune radius. As remarked by \cite{Davis2017}, some practical orbits exist within the Earth-Moon NRHOs. A 9:2 resonance with the Moon synodic period ($\sim$29.5 days) can be found at an altitude of $\sim$1500 km, whereas another 4:1 resonance arises at $\sim$4150 km, which are useful to avoid Earth eclipses at all times.  

\subsection{Relative motion in the R3BP}

For relative dynamics, following \cite{Franzini2017}, a local frame (LVLH) is employed. The frame is denoted by $L:\{\mathbf{r}_t, \mathbf{i}_L, \mathbf{j}_L, \mathbf{k}_L\}$, where $\mathbf{r}_t$ is the target position,  $\mathbf{k}_L$ is pointing towards the second primary, $\mathbf{j}_L$ is in the opposite direction to the target kinetic momentum as view from the $S$ frame with respect to the second primary and $\mathbf{i}_L$ completes the right-handed frame. Fig.\ref{fig:LVLH_frame} shows the $L$ frame as well as the target position $\mathbf{r}_t$, the chaser position $\mathbf{r}$ and the relative position $\pmb{\rho}=\mathbf{r}-\mathbf{r}_t$. The relative  dynamics in the $L$ frame is given by 
$
\left.\pmb{\ddot{\rho}}\right|_L=\left.\mathbf{\ddot{r}}\right|_L-\left.\mathbf{\ddot{r}}_t\right|_L,$
 which can be further developed by using Eq.\eqref{eq:R3BP_dyn_synodic}, reaching
\begin{equation}
\begin{aligned}
\left.\pmb{\ddot{\rho}}\right|_L=&-2\pmb{\omega}_{L/I}\times\left.\pmb{\dot{\rho}}\right|_L-\pmb{\omega}_{L/I}\times(\pmb{\omega}_{L/I} \times\pmb{\rho})\\     &-\left.\pmb{\dot{\omega}}_{L/I}\right|_L\times\pmb{\rho}-\mu_1\left(\frac{\pmb{\rho}+\mathbf{r}_{1t}}{\|\pmb{\rho}+\mathbf{r}_{1t}\|^3_2}-\frac{\mathbf{r}_{1t}}{\|\mathbf{r}_{1t}\|^3_2}\right)\\
&-\mu_2\left(\frac{\pmb{\rho}+\mathbf{r}_{2t}}{\|\pmb{\rho}+\mathbf{r}_{2t}\|^3_2}-\frac{\mathbf{r}_{2t}}{\|\mathbf{r}_{2t}\|^3_2}\right)+\mathbf{u},
\end{aligned} \label{eq:R3BP_rel_dyn_LVLH}
\end{equation} 
where $\mathbf{r}_{1t}=\mathbf{r}_t-\mathbf{r}_1$ and $\mathbf{r}_{2t}=\mathbf{r}_t-\mathbf{r}_2$ denote the relative position of the target with respect to the primaries. Note
\begin{align}
\pmb{\omega}_{L/I}&=\pmb{\omega}_{L/S}+\pmb{\omega}_{S/I},\\
\left.\pmb{\dot{\omega}}_{L/I}\right|_L&=\left.\pmb{\dot{\omega}}_{L/S}\right|_L+\left.\pmb{\dot{\omega}}_{S/I}\right|_S-\pmb{\omega}_{L/S}\times \pmb{\omega}_{S/I},
\end{align}
hence, $\mathbf{r}_t$, $\pmb{\omega}_{L/S}$ and $\left.\pmb{\dot{\omega}}_{L/S}\right|_L$ depend on the target motion with respect to the synodic frame whereas $\mathbf{r}_1$, $\mathbf{r}_2$, $\pmb{\omega}_{S/I}$ and $\left.\pmb{\dot{\omega}}_{S/I}\right|_S$ depend on the primaries motion (Eq.\eqref{eq:R3BP_rel_dyn_LVLH} is still valid if the primaries  evolve in elliptic orbits).

\subsection{Linearized relative motion in the R3BP}

Considering close-range rendezvous operations, that is, $\|\mathbf{r}_{1t}\|_2, \|\mathbf{r}_{2t}\|_2 \gg \|\pmb{\rho}\|_2$, one has
\begin{equation}
\frac{\mathbf{r}}{\|\mathbf{r}\|^3_2} \approx \frac{\mathbf{r}_0}{\|\mathbf{r}_0\|^3_2}-\frac{1}{\|\mathbf{r}_0\|^3_2}\left(\mathbf{I}-3\frac{\mathbf{r}_0\mathbf{r}_0^T}{\|\mathbf{r}_0\|^2_2}\right)(\mathbf{r}-\mathbf{r}_0), \label{eq:lin_gravity_example}
\end{equation}
being $\mathbf{r}_0$ the linearization point. Introducing the linearization of Eq.\eqref{eq:lin_gravity_example} into Eq.\eqref{eq:R3BP_rel_dyn_LVLH}, one obtains
\begin{equation}
\begin{aligned}
\pmb{\ddot{\rho}}=&-\left(\pmb{\dot{\Omega}}_{L/I}+\pmb{\Omega}^2_{L/I}-\frac{\mu_1}{r_{1t}^3}\left(\mathbf{I}-3\frac{\mathbf{r}_{1t}\mathbf{r}_{1t}^T}{r_{1t}^2}\right)\right. \left.-\frac{\mu_2}{r_{2t}^3}\left(\mathbf{I}-3\frac{\mathbf{r}_{2t}\mathbf{r}_{2t}^T}{r_{2t}^2}\right) \right)\pmb{\rho}-2\pmb{\Omega}_{L/I}\pmb{\dot{\rho}}+\mathbf{u},
\end{aligned}
\label{eq:R3BP_lin_rel_dyn_LVLH}
\end{equation}
where $\pmb{\Omega}_{L/I}$ and $\pmb{\dot{\Omega}}_{L/I}$ are the cross-product matrices  associated to $\pmb{\omega}_{L/I}$ and $\dot{\pmb{\omega}}_{L/I}$ respectively, see \cite{Wie2008}. This can be written as a  linear time-varying (LTV) system:
\begin{equation}
\frac{d}{dt}
\begin{bmatrix}
\pmb{\rho}\\
\pmb{\dot{\rho}}
\end{bmatrix}=
\begin{bmatrix}
\mathbf{0} & \mathbf{I}\\
\mathbf{A}_{\pmb{\dot{\rho}}\pmb{\rho}} & -2\pmb{\Omega}_{L/I}
\end{bmatrix}
\begin{bmatrix}
\pmb{\rho}\\
\pmb{\dot{\rho}}
\end{bmatrix}
+
\begin{bmatrix}
\mathbf{0}\\
\mathbf{I}\\
\end{bmatrix}
\mathbf{u},
\label{eq:R3BP_lin_rel_dyn_LVLH_matrix}
\end{equation}
where
\begin{equation}
\begin{aligned}
\mathbf{A}_{\pmb{\dot{\rho}}\pmb{\rho}}=&-\left(\pmb{\dot{\Omega}}_{L/I}+\pmb{\Omega}^2_{L/I}-\frac{\mu_1}{r_{1t}^3}\left(\mathbf{I}-3\frac{\mathbf{r}_{1t}\mathbf{r}_{1t}^T}{r_{1t}^2}\right)\right.
\left.-\frac{\mu_2}{r_{2t}^3}\left(\mathbf{I}-3\frac{\mathbf{r}_{2t}\mathbf{r}_{2t}^T}{r_{2t}^2}\right) \right).
\end{aligned}
\end{equation}
Defining $\mathbf{x}=[\pmb{\rho}^T, \pmb{\dot{\rho}}^T]^T$, Eq.\eqref{eq:R3BP_lin_rel_dyn_LVLH_matrix} is of the form $\mathbf{\dot{x}}(t)=\mathbf{A}(t)\mathbf{x}(t)+\mathbf{B}\mathbf{u}(t)$, which has as general solution, see \cite{Kamen2010}, 
\begin{equation}
\mathbf{x}(t)=\pmb{\phi}(t,t_0)\mathbf{x}_0+\int^t_{t_0}\pmb{\phi}(t,\tau)\mathbf{B}\mathbf{u}(\tau)d\tau, \label{eq:LTV_system_general_solution}
\end{equation}
with $\pmb{\phi}(t,t_0)$  the state transition matrix, verifying
\begin{equation}
\pmb{\dot{\phi}}(t,t_0)=\mathbf{A}(t)\pmb{\phi}(t,t_0), \quad \pmb{\phi}(t_0,t_0)=\mathbf{I}. \label{eq:LTV_transition_matrix}
\end{equation} 
\section{Rendezvous planning problem}

Next, the control inputs are described and parameterized;  then, the objective function and the constraints are described. Finally, the rendezvous problem is stated.   

\subsection{Control input}

In this work, both chemical and electric thrusters are considered; thus, 
$\mathbf{u}=\mathbf{u}_C+\mathbf{u}_E$,
where $\mathbf{u}_C$ and $\mathbf{u}_E$ denote the chemical and electric accelerations respectively. For the chemical thrusters, the control signal can be described by impulses (i.e. instantaneous changes of velocity) 
\begin{equation}
\lim_{\Delta t\to0}\int^{t_k+\Delta t}_{t_k}\mathbf{u}_C(t)dt=\Delta \mathbf{V}(t_k)\delta(t-t_k), \label{eq:impulse_control_action}
\end{equation}
where $t_k$ is the impulse application time. On the other hand, electric thrusters provide continuous thrust and are assumed to depend linearly on some parameters $\pmb{\xi}\in\mathbb{R}^{3n_{\xi}}$
\begin{equation}
\mathbf{u}_E(t)=\mathbf{B}_{\xi}(t)\pmb{\xi}, \quad \pmb{\xi}=[\pmb{\xi}_1^T, \pmb{\xi}_2^T \hdots \pmb{\xi}_{n_{\xi}}^T]^T, \label{eq:continuous_control_action}
\end{equation}
where the matrix $\mathbf{B}_{\xi} \in \mathbb{R}^{3\times3n_{\xi}}$, following \cite{Sanchez2020}, is described by B-splines, (see \cite{Kress1998} for more details about them).  Thus
\begin{equation}
\mathbf{u}_E(t)=\sum^{n_c}_{j=1}B_{j,q}(t)\pmb{\xi}_j,
\end{equation}
where $B_{j,q}$ are $q$th order B-splines built on the knots sequence $\mathbf{t}_{\text{knots}} \in \mathbb{R}^{n_{\text{knots}}}$ while $\pmb{\xi}_j \in \mathbb{R}^3$ are the control points.  If none of the internal knots is repeated, the B-splines intrinsically assure continuity up to $C^q$. Given the order $q$ and the number of coefficients $n_c$, the number of knots must satisfy $n_{\text{knots}}=n_c+q+1$.

\subsection{Objective function}

The chosen objective function seeks to minimize the control effort of both the chemical and electric thrusters
\begin{equation}
J = \beta\sum^{N}_{k=0}\lVert\Delta \mathbf{V}(t_k)\rVert^2_2+(1-\beta)\lVert\pmb{\xi}\rVert^2_2, \label{eq:obj_func}
\end{equation}
where $N+1$ is the number of impulses along the manoeuvre and $\beta$ is a weight parameter. 

\subsection{Constraints}

Three types of constraints are considered in this paper. Firstly, path constraints on the relative state; secondly the control signals are bounded; and finally, initial and terminal states values are prescribed.  

\subsubsection{Path constraints}

\begin{figure}[h] 
	\begin{center}
		\includegraphics[width=6.5cm,height=6.5cm,keepaspectratio]{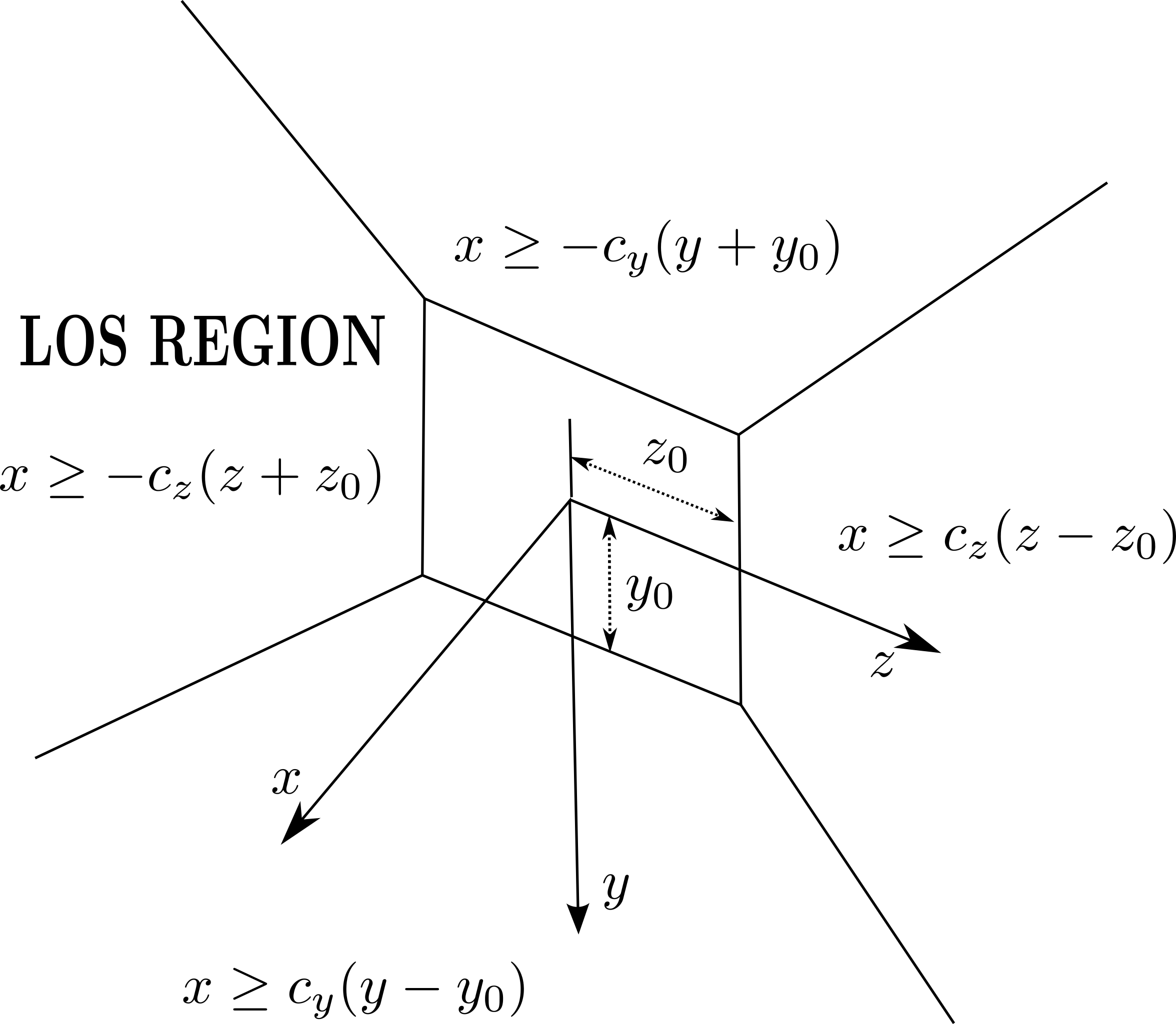}
	\end{center}
	\caption{LOS region}	
	\label{fig:LOS_region}
\end{figure}

For sensing purposes, it is required that the chaser is at all time visible from the docking port, see \cite{Breger2008}. This constitutes the line-of-sight (LOS) constraint. The LOS region can be defined by the equations $x\geq c_y(y-y_0)$, $x\geq-c_y(y+y_0)$, $x\geq c_z(z-z_0)$, $x\geq -c_z(z+z_0)$ and $x\geq 0$; these equations limit the relative translational state space by five planes as shown in Fig.\ref{fig:LOS_region}. One can define the LOS algebraically, at any instant $t$ as $\mathbf{A}_L\mathbf{x}(t)\leq \mathbf{b}_L$, where
\begin{equation}
\mathbf{A}_L=
\begin{bmatrix}
-1 & c_y & 0 & 0 & 0 & 0\\
-1 & -c_y & 0 & 0 & 0 & 0\\
-1 & 0 & c_z & 0 & 0 & 0\\
-1 & 0 & -c_z & 0 & 0 & 0\\
-1 & 0 & 0 & 0 & 0 & 0\\
\end{bmatrix}, \> \>
\mathbf{b}_L=
\begin{bmatrix}
c_y y_0\\
c_y y_0\\
c_z z_0\\
c_z z_0\\
0
\end{bmatrix}.\label{eq:LOS_constraint}
\end{equation}

\subsubsection{Control bounds}

For each type of thruster and each direction, it is assumed that its control action is bounded above and below in the same way 
\begin{align}
-\Delta \mathbf{V}_{\text{max}} &\leq \Delta\mathbf{V}(t) \leq \Delta \mathbf{V}_{\text{max}},\\
-\mathbf{u}_{\text{max}} &\leq \mathbf{B}_{\xi}(t)\pmb{\xi} \leq \mathbf{u}_{\text{max}}.
\end{align}

\subsubsection{Boundary constraints}

The chaser departs from a given location and velocity at the initial time $t_0$ and has to met the target at the end of the manoeuvre $t_f$ 
\begin{equation}
\mathbf{x}(t_0)=\mathbf{x}_0, \quad \mathbf{x}(t_f)=\mathbf{0}. \label{eq:boundary_constraints}
\end{equation}

\subsection{Rendezvous problem}

Putting together the objective function given by Eq.\eqref{eq:obj_func}, the constraints of Eq.\eqref{eq:LOS_constraint}-\eqref{eq:boundary_constraints} and inserting the control inputs expressions of Eq.\eqref{eq:impulse_control_action}-\eqref{eq:continuous_control_action} into Eq.\eqref{eq:LTV_system_general_solution}, one obtains the planning rendezvous problem
\begin{equation}
\begin{aligned}
& \underset{\Delta \mathbf{V}, \pmb{\xi}}{\text{min}}
& & \beta\sum^{N}_{k=0}\lVert\Delta \mathbf{V}(t_k)\rVert^2_2+(1-\beta)\lVert\pmb{\xi}\rVert^2_2, \\
& \text{s.t.}
& & \mathbf{x}(t)=\pmb{\phi}(t,t_0)\mathbf{x}_0+\int^{t}_{t_0}\pmb{\phi}(t,\tau)\mathbf{B}\mathbf{B}_{\xi}(\tau)\pmb{\xi}d\tau+\sum^N_{k=1}\pmb{\phi}(t,t_k)\mathbf{B}\Delta \mathbf{V}(t_k)\delta(t-t_k),\\
&&& \mathbf{A}_L\mathbf{x}(t)\leq \mathbf{b}_L,\\
&&& -\Delta\mathbf{V}_{\text{max}} \leq \Delta \mathbf{V}(t) \leq \Delta\mathbf{V}_{\text{max}},\\
&&& -\mathbf{u}_{\text{max}} \leq \mathbf{B}_{\xi}(t)\pmb{\xi} \leq \mathbf{u}_{\text{max}}, \\
&&& \mathbf{x}(t_0)=\mathbf{x}_0,\\
&&& \mathbf{x}(t_f)=\mathbf{0}.\\
\end{aligned}\label{eq:opt_orig}
\end{equation}
Note that the optimization problem \eqref{eq:opt_orig} has a quadratic objective function and linear constraints.

\section{Robust MPC formulation}

In this section, a robust MPC scheme, in the spirit of the chance constrained approach (see \cite{Gavilan2012}), is formulated; firstly the problem is discretized and disturbances are included into the model. Secondly, it is shown how to robustify the controller to tackle these disturbances in a probabilistic way. Finally, a disturbance estimator, to compute on-line the perturbations statistical properties, is developed.

\subsection{Discretized prediction of the state}

To transform the rendezvous problem \eqref{eq:opt_orig} into a finite tractable program, the relative dynamics is discretized with respect to time. In particular, the manoeuvre duration is divided into $N$ equally distributed sampling times $\Delta T=(t_f-t_0)/N$ resulting into $N+1$ time nodes. Denote by $\mathbf{x}_{k+j}$ the state at the instant $t_{k+j}$ when an impulse $\Delta \mathbf{V}_{k+j}$ is applied. The discrete propagation from the instant $t_k$ to $t_{k+j}$ is given by
\begin{equation}
\begin{aligned}
&\mathbf{x}_{k+j}=\pmb{\phi}(t_{k+j},t_k)\mathbf{x}_k+\sum^{j}_{i=0}\pmb{\phi}(t_{k+j},t_{k+i})\mathbf{B}\Delta\mathbf{V}_{k+i}
+\sum^{j-1}_{i=0}\pmb{\phi}(t_{k+j},t_{k+i})
\\
& \times\left(\int^{t_{k+j+1}}_{t_{k+j}}\pmb{\phi}(t,t_{k+j})\mathbf{B}\mathbf{B}_{\xi}(t)dt\right)\pmb{\xi}+\sum^{j}_{i=0}\pmb{\phi}(t_{k+j},t_{k+i})\pmb{\delta}_{k+i}, \quad t_{k+j}=(k+j)\Delta T, \label{eq:state_prop_robust_dV}
\end{aligned}
\end{equation}
where an additive disturbance to the state, denoted by $\pmb{\delta}$, is added at each node $t_{k+j},~j=0\hdots N$. The term $\pmb{\delta}$ could model navigation errors, as a position disturbance, and perturbation forces, as a velocity disturbance. Note that $N+1$ impulses are considered to be applied at the nodes $t_{k+j}$ along the manoeuvre. To ease the notation, following \cite{Vazquez2017}, a compact formulation is developed. Defining the following stack vectors, $\mathbf{x}_S \in \mathbb{R}^{6(N+1)}$, $\Delta\mathbf{V}_S \in \mathbb{R}^{3(N+1)}$, $\pmb{\xi}_S\in\mathbb{R}^{3n_{\xi}}$ and $\pmb{\delta}_S \in \mathbb{R}^{3(N+1)}$, 
\begin{equation*}
\begin{aligned}
\mathbf{x}_S(k)&=\begin{bmatrix}
\mathbf{x}^T_{k}, & \mathbf{x}^T_{k+1} & \hdots & \mathbf{x}^T_{k+N}\end{bmatrix}^T,\\
\Delta\mathbf{V}_S(k)&=\begin{bmatrix}
\Delta\mathbf{V}^T_{k}, & \Delta\mathbf{V}^T_{k+1} & \hdots & \Delta\mathbf{V}^T_{k+N}\end{bmatrix}^T,\\
\pmb{\xi}_S(k)&=\begin{bmatrix}
\pmb{\xi}^T_{k+1}, & \pmb{\xi}^T_{k+2} & \hdots & \pmb{\xi}^T_{k+n_{\xi}}
\end{bmatrix}^T,\\
\pmb{\delta}_S(k)&=
\begin{bmatrix}
\pmb{\delta}^T_{k}, & \pmb{\delta}^T_{k+1} & \hdots & \pmb{\delta}^T_{k+N}\\
\end{bmatrix}^T,
\end{aligned} \label{eq:stack_vectors}
\end{equation*}
and the stack matrices $\mathbf{F}$, $\mathbf{G}_{\Delta V}$, $\mathbf{G}_{\xi}$ and  $\mathbf{G}_{\delta}$
\begin{equation*}
\begin{aligned}
\mathbf{F}_k&=\begin{bmatrix}
\mathbf{I}, & \pmb{\phi}^T(t_{k+1},t_k) & \hdots & \pmb{\phi}^T(t_{k+N},t_k)\\
\end{bmatrix}^T,\\
\mathbf{G}_{k,\delta}&=
\begin{bmatrix}
\mathbf{I} & \mathbf{0}_{6\times6} & \hdots & \mathbf{0}_{6\times6}\\
\pmb{\phi}(t_{k+1},t_{k}) & \mathbf{I} & \hdots & \mathbf{0}_{6\times6}\\
\vdots & \vdots & \ddots & \vdots\\
\pmb{\phi}(t_{k+N},t_{k}) & \pmb{\phi}(t_{k+N-1},t_{k}) & \hdots & \mathbf{I}\\
\end{bmatrix}, 
\end{aligned}\label{eq:stack_matrices_1}
\end{equation*}
\begin{equation*}
\mathbf{G}_{k,\Delta V}=
\begin{bmatrix}
\mathbf{B} & \mathbf{0}_{6\times3} & \hdots & \mathbf{0}_{6\times3}\\
\pmb{\phi}(t_{k+1},t_{k})\mathbf{B} & \mathbf{B} & \hdots & \mathbf{0}_{6\times3}\\
\vdots & \vdots & \ddots & \vdots\\
\pmb{\phi}(t_{k+N},t_{k})\mathbf{B} & \pmb{\phi}(t_{k+N-1},t_{k})\mathbf{B} & \hdots & \mathbf{B}\\
\end{bmatrix}, \label{eq:stack_matrices_2}
\end{equation*}

\begin{equation*}
\mathbf{G}_{k,\xi} =
\begin{bmatrix}
\mathbf{0}_{6\times3} & \hdots & \mathbf{0}_{6\times3}\\
\mathbf{B}_{u_{k+1}}(t_{k+1}) & \hdots & \mathbf{B}_{u_{k+n_{\xi}}}(t_{k+1})\\
\vdots  & \ddots & \vdots \\
\sum^{N}_{i=1}\pmb{\phi}(t_{k+N},t_{k+i})\mathbf{B}_{u_{k+1}}(t_{k+i}) & \hdots & \sum^{N}_{i=1}\pmb{\phi}(t_{k+N},t_{k+i})\mathbf{B}_{u_{k+n_{\xi}}}(t_{k+i})\\
\end{bmatrix},
\end{equation*}

where $\mathbf{0}$ denotes a matrix full of zeros and
\begin{equation}
\mathbf{B}_{u_{k+l}}(t_{k+j})=\int^{t_{k+j}}_{t_{k+j-1}}\pmb{\phi}(t,t_{k+j-1})\mathbf{B}\mathbf{B}_{k+l,\xi}(t)dt,
\end{equation}
being $\mathbf{B}_{l,\xi}\in\mathbb{R}^{3\times3}$ the diagonal submatrices of $\mathbf{B}_{\xi}$. Then, 
\begin{equation}
\mathbf{x}_S(k)=\mathbf{F}_k\mathbf{x}_k+\mathbf{G}_{k,\Delta V}\Delta\mathbf{V}_S(k)+\mathbf{G}_{k,\xi}\pmb{\xi}_S(k)+\mathbf{G}_{k,\delta}\pmb{\delta}_S(k). \label{eq:state_prop_robust_compact}
\end{equation}

\subsection{Objective function and constraints}

For the robust controller, a terminal penalty term is added instead of the terminal constraint \eqref{eq:boundary_constraints}, which is removed. This constraint relaxation prevents the optimization to become infeasible, see \cite{Chen1998}, and can potentially improve asymptotic stability properties, see \cite{Limon2006}. Due to the disturbance terms added to the state propagation, see Eq.\eqref{eq:state_prop_robust_compact}, the state $\mathbf{x}$ evolves stochastically. Therefore, mathematical expectation of the state can be taken as $\mathbf{\hat{x}}_{k+j|k}=E[\mathbf{x}_{k+j}]$, given $\mathbf{x}_k$ and $\mathbf{\hat{x}}_S(k+j|k)=E[\mathbf{x}_S(k+j)]$. 

The robust objective function becomes
\begin{equation}
\begin{aligned}
&J(k)=\sum^N_{j=0}\mathbf{\hat{x}}_{k+j|k}^T\mathbf{R}(k+j)\mathbf{\hat{x}}_{k+j|k}+\beta\sum^N_{j=0}\Delta\mathbf{V}^T_{k+j}\mathbf{I}\Delta\mathbf{V}_{k+j}+(1-\beta)\sum^{n_{\xi}}_{l=0}\pmb{\xi}^T_{k+l}\mathbf{I}\pmb{\xi}_{k+l}, \label{eq:obj_func_mean_state}
\end{aligned}
\end{equation}
where the terminal weight matrix $\mathbf{R}$ is defined as in \cite{Gavilan2012}
\begin{equation}
\mathbf{R}(k+j)=\gamma h(k+j-k_a)
\begin{bmatrix}
\mathbf{I} & \mathbf{0}_{3 \times 3}\\
\mathbf{0}_{3 \times 3} & \mathbf{0}_{3 \times 3}\\
\end{bmatrix},
\end{equation}
being $h$ the step function, $k_a$ the desired arrival time and $\gamma$ a large positive number. The instants $k+j>k_a$ are weighted because it is desired to arrive at the target at the instant $t_{k_a}$ and remain there. Defining $E[\pmb{\delta}_S(k+j)]=\pmb{\bar{\delta}}_S$, the robust objective function of Eq.\eqref{eq:obj_func_mean_state} is expressed compactly as
\begin{equation}
\begin{aligned}
J(k)=&\hat{\mathbf{x}}^T_S(k)\mathbf{R}_S\hat{\mathbf{x}}_S(k)+\Delta\mathbf{V}^T_S(k)\mathbf{Q}_{\Delta V}\Delta\mathbf{V}_S(k)+\pmb{\xi}^T_S(k)\mathbf{Q}_{\xi}\pmb{\xi}_S(k).
\end{aligned}
\end{equation}
On the other hand, $\mathbf{Q}_{\Delta V}=\beta\mathbf{I}$, $\mathbf{Q}_{\xi}=(1-\beta)\mathbf{I}$ and the matrix $\mathbf{R}_S$ is
\begin{equation}
\mathbf{R}_S=
\begin{bmatrix}
\mathbf{R}(k) & \hdots & \mathbf{0}_{6\times6}\\
\vdots & \ddots & \vdots\\
\mathbf{0}_{6\times6} & \hdots & \mathbf{R}(k+N)\\
\end{bmatrix}.
\end{equation}

The LOS constraint, given by Eq.\eqref{eq:LOS_constraint}, is expressed as
\begin{equation}
\mathbf{A}_{LS}\mathbf{x}_S(k)\leq \mathbf{b}_{LS},
\label{eq:LOS_constraint_disturbance}
\end{equation}
where $\mathbf{A}_{LS}\in\mathbb{R}^{5N\times6(N+1)}$ and $\mathbf{b}_{LS} \in \mathbb{R}^{5N}$ stack the LOS constraint matrix and vector of Eq.\eqref{eq:LOS_constraint} respectively. Similarly, the chemical thrusters bounds can be written as
\begin{equation}
-\Delta \mathbf{V}_{S,\text{max}}\leq \Delta\mathbf{V}_S(k) \leq \Delta \mathbf{V}_{S,\text{max}}.
\end{equation}
 The electric thrusters constraint is tackled discretely:
\begin{equation}
-\mathbf{u}_{S,\text{max}}\leq \mathbf{B}_{k,S_\xi}\pmb{\xi}_S(k) \leq \mathbf{u}_{S,\text{max}},
\end{equation}
where
\begin{equation}
\mathbf{B}_{k,S_\xi}=\begin{bmatrix}
\mathbf{B}^T_{\xi}(t_k), & \mathbf{B}^T_{\xi}(t_{k+1})&\hdots&\mathbf{B}^T_{\xi}(t_{k+n_u})\\
\end{bmatrix}^T,
\end{equation}
with $n_u+1$ instants equispaced by $\Delta T_{n_u}=(t_{k+N}-t_k)/n_u$. 

\subsection{Robust satisfaction of constraints} \label{subsec:robust_constraints}

Assuming that $\pmb{\delta}_S$ is a random term with unknown bounds, the LOS inequality of Eq.\eqref{eq:LOS_constraint_disturbance} is made to be satisfied with a certain probability (chance-constrained). Introducing the bounding term $\mathbf{b}_{\delta}(k)$ into Eq.\eqref{eq:LOS_constraint_disturbance}
\begin{equation}
\begin{aligned}
\mathbf{A}_{LS}(\mathbf{G}_{k,\Delta V}\Delta\mathbf{V}_S(k)+\mathbf{G}_{k,\xi}\pmb{\xi}_S(k))&\leq \mathbf{b}_{LS}-\mathbf{A}_{LS}\mathbf{F}_k\mathbf{x}_k+\mathbf{b}_{\delta}(k)\\
&\leq
\mathbf{b}_{LS}-\mathbf{A}_{LS}(\mathbf{F}_k\mathbf{x}_k+\mathbf{G}_{k,\delta}\pmb{\delta}_S(k)).
\end{aligned}
\label{eq:LOS_constraint_disturbance_bounded}
\end{equation}
The probability of constraint satisfaction, by adding the bounding term $\mathbf{b}_{\delta}$, should be near one. This guarantees that the chaser remains within the LOS region for almost all perturbations. Considering that the disturbances are normally distributed, $\pmb{\delta}\sim N_6(\pmb{\bar{\delta}},\pmb{\Sigma}_{\delta})$, with known mean, $\pmb{\bar{\delta}}$, and covariance matrix, $\pmb{\Sigma}_{\delta}=\pmb{\Sigma}^T_{\delta}\succ0$, the following relation holds (see \cite{Rencher1998} for more details)
\begin{equation}
\pmb{\delta}\sim N_6(\pmb{\bar{\delta}}, \pmb{\Sigma}_{\delta}) \longrightarrow (\pmb{\delta}-\pmb{\bar{\delta}})^T\pmb{\Sigma}^{-1}_{\delta}(\pmb{\delta}-\pmb{\bar{\delta}})\sim \chi^2(6) \label{eq:disturbance_chi},
\end{equation} 
where $\chi^2(6)$ is a chi-square probability distribution with six degrees of freedom. Making the hypothesis that the statistical properties of the disturbances are time-invariant (quasi-steady approach), Eq.\eqref{eq:disturbance_chi} is valid at all times
\begin{equation}
(\pmb{\delta}_{k+j}-\pmb{\bar{\delta}})^T\pmb{\Sigma}^{-1}_{\delta}(\pmb{\delta}_{k+j}-\pmb{\bar{\delta}})\sim \chi^2(6), \quad j=0\hdots N,
\end{equation}
hence the following probabilistic relations hold
\begin{equation*}
\text{P}(\chi^2(6) \leq \alpha)=p \longrightarrow (\pmb{\delta}_{k+j}-\pmb{\bar{\delta}})^T\pmb{\Sigma}^{-1}_{\delta}(\pmb{\delta}_{k+j}-\pmb{\bar{\delta}})\leq \alpha,
\end{equation*}
where finding $\alpha$ from a given $p$, the right side inequality is guaranteed with probability $p$. Then, the parameter $p$ is the probability of constraint satisfaction and should be as close to unity as possible. The bounding term $\mathbf{b}_{\delta}(k)$ can be found by solving the following minimization problem for each row $i$ of $-\mathbf{A}_{LS}\mathbf{G}_{k,\delta}$ denoted as $\mathbf{a}_i(k)$
\begin{equation}
\begin{aligned}
& \underset{\pmb{\delta}_S}{\text{min}}
& & (\mathbf{b}_\delta(k))_i=\mathbf{a}_i(k)\pmb{\delta}_S, \\
& \text{s.t.}
& & (\pmb{\delta}_{k+j}-\pmb{\bar{\delta}})^T(\alpha \pmb{\Sigma}_{\delta})^{-1}(\pmb{\delta}_{k+j}-\pmb{\bar{\delta}})\leq 1,\\
\end{aligned}\label{eq:LOS_bound_opt_1}
\end{equation}
It can be proved, see \cite{Gavilan2012}, that the rows of $\mathbf{b}_{\delta}(k)$ are
\begin{equation}
(\mathbf{b}_{\delta}(k))_i=\sum^{N}_{j=0}\left(-\sqrt{\mathbf{a}_{ij}\mathbf{H}^{-1}\mathbf{a}_{ij}}+\mathbf{a}_{ij}\pmb{\bar{\delta}} \right).  \label{eq:LOS_bound_b}
\end{equation}
Once the vector $\mathbf{b}_{\delta}(k)$ is computed through Eq.\eqref{eq:LOS_bound_b}, the control input at time $t_k$ is obtained by solving the following robust program
\begin{equation}
\begin{aligned}
& \underset{\Delta\mathbf{V}_S(k),\pmb{\xi}_S(k)}{\text{min}}
& & J(\mathbf{x}_k, \Delta\mathbf{V}_S(k), \pmb{\xi}_S(k), \pmb{\bar{\delta}}_S(k)), \\
& \text{s.t.}
& & \mathbf{A}_{LS}(\mathbf{G}_{k,\Delta V}\Delta\mathbf{V}_S(k)+\mathbf{G}_{k,\xi}\pmb{\xi}_S(k))\leq\mathbf{b}_{LS}-\mathbf{A}_{LS}\mathbf{F}_k\mathbf{x}_k+\mathbf{b}_{\delta}(k),\\
&&& -\Delta\mathbf{V}_{S,\text{max}}\leq\Delta\mathbf{V}_S(k)\leq\Delta\mathbf{V}_{S,\text{max}},\\
&&& -\mathbf{u}_{S,\text{max}}\leq\mathbf{B}_{k,S_\xi}\pmb{\xi}_S(k)\leq\mathbf{u}_{S,\text{max}},
\end{aligned}\label{eq:opt_robust_MPC}
\end{equation}
which is a quadratic programming (QP)  problem.

\subsection{Disturbance estimator}

The robust satisfaction of constraints, presented in the section \ref{subsec:robust_constraints}, requires a priori knowledge of the perturbations statistical properties, $\pmb{\bar{\delta}}$ and $\pmb{\Sigma}_{\delta}$. However, such properties are typically unknown and they have to be estimated on-line. Since the disturbances have been assumed as normally distributed such that $\pmb{\delta}\sim N_6(\pmb{\bar{\delta}},\pmb{\Sigma}_{\delta})$, the normal distribution parameters $\pmb{\bar{\delta}}$ and $\pmb{\Sigma}_{\delta}$ are estimated a posteriori at each time $k$ by taking into account all past disturbances
\begin{equation}
\pmb{\delta}_i=\mathbf{x}_{i+1}-\pmb{\phi}(t_{i+1},t_{i})\mathbf{x}_i-\int^{t_{i+1}}_{t_i}\pmb{\phi}(t_{i+1},\tau)\mathbf{B}\mathbf{u}(\tau)d\tau,
\end{equation}
with $i=1\hdots k-1$. The estimates of $\pmb{\bar{\delta}}$ and $\pmb{\Sigma}_{\delta}$ at time $k$, based on disturbances up to $k-1$, are named as $\pmb{\hat{\delta}}_k$ and $\pmb{\hat{\Sigma}}_{k,\delta}$, and following \cite{Gavilan2012} one can use 
%
 recursive formulas for their estimation as follows
\begin{eqnarray*}
\pmb{\hat{\delta}}_k&=&
 \frac{e^{-\lambda}}{\gamma_k}(\gamma_{k-1}\pmb{\hat{\delta}}_{k-1}+\pmb{\delta}_{k-1}),\\
\pmb{\hat{\Sigma}}_{k,\delta}&=
&\frac{e^{-\lambda}}{\gamma_k}\left(\gamma_{k-1}\pmb{\hat{\Sigma}}_{k-1}+(\pmb{\delta}_{k-1}-\pmb{\hat{\delta}}_k)(\pmb{\delta}_{k-1}-\pmb{\hat{\delta}}_k)^T\right),
\end{eqnarray*}
with $\pmb{\hat{\delta}}_0=\mathbf{0}$ and $\pmb{\hat{\Sigma}}_{0,\delta}=\mathbf{0}$.

\section{Results} 

In this section, an application case of rendezvous with a target located in an Earth-Moon NRHO is considered. A comparison between the proposed chance-constrained MPC algorithm against a non-robust MPC is carried out.

\subsection{Simulation model}

The non-linear R3BP relative dynamics given by Eq.\eqref{eq:R3BP_rel_dyn_LVLH} are used to obtain the numerical results of this section. As reported in \cite{Franzini2017}, the position error between the linear and non-linear models increases faster at the NRHO perilune ($\sim40~\text{m}$ in $1~\text{h}$) compared to its apolune ($\sim2~\text{m}$ in $1~\text{h}$). The minimum and maximum distance between Earth and Moon are taken as $\|\mathbf{\underline{r}}_{12}\|=363104~\text{km}$ and $\|\mathbf{\overline{r}}_{12}\|=405696$ km, whereas the primaries gravitational parameters are $\mu_1=398600.4~\text{km}^3/\text{s}^2$ and $\mu_2=4904.869~\text{km}^3/\text{s}^2$. The manoeuvre is considered to take place when the distance between Moon and Earth is minimal.

Apart from model mismatch, numerical integration is required to obtain the LTV transition matrix with Eq.\eqref{eq:LTV_transition_matrix}, hence cumulative integration errors are expected to arise. Another source of disturbances is the computation of the target NRHO which is done with the continuation software AUTO (see \cite{AUTO2012}), using the CR3BP model, see Eq.\eqref{eq:CR3BP_dyn_synodic}. The target $L_2$ Southern NRHO is taken as the one with $\nu=1.0120$, $T=10.35~\text{days}$ and closest distance to the Moon surface of $15674~ \text{km}$, see Fig.\ref{fig:NRHO_stability_periods}.

Regarding the thrusters performance, in the same sense as \cite{Gavilan2012}, the real control inputs $\Delta\mathbf{V}_{\text{real}}=[\Delta V_x, \Delta V_y, \Delta V_z]^T$ and $\mathbf{u}_{\text{real}}=[u_x, u_y, u_z]^T$ do not match the computed control signals $\Delta\mathbf{V}$ and $\mathbf{u}_E$
\begin{align}
\Delta\mathbf{V}_{\text{real}}&=\mathbf{R}(\pmb{\delta \theta})(\Delta\mathbf{V}+\pmb{\delta}\mathbf{V}), \label{eq:perturbations_dV}\\
\mathbf{u}_{\text{real}}&=\mathbf{R}(\pmb{\delta \theta})(\mathbf{u}_E+\pmb{\delta}\mathbf{u}_E), \label{eq:perturbations_continuous}
\end{align}
where $\mathbf{R}$ is a rotation matrix and $\pmb{\delta \theta}\sim N_3(\pmb{\delta} \pmb{\bar{\theta}}, \pmb{\Sigma}_{\delta \theta})$ is a vector of small random angles modelling imperfect alignment of thrusters, whereas $\pmb{\delta}\mathbf{V} \sim N_3(\pmb{\delta}\mathbf{\bar{V}}, \pmb{\Sigma}_{\delta V})$ and $\pmb{\delta}\mathbf{u}_E \sim N_3(\pmb{\delta}\bar{\mathbf{u}}_E, \pmb{\Sigma}_{\delta u_E})$ are additive random noises to the impulse or electric thrust amplitude respectively. Note that $N_n$ denotes a $n$-dimensional gaussian distribution. 

\subsection{Simulation results}

In this section, the previously designed robust controller performance is evaluated for each one of the thrusters configurations. The initial manoeuvre time is chosen at the instant when the target is closest to the Moon (perilune), thus potentially representing a lunar sample return scenario, see \cite{Landgraf2018}.

The simulations are done in MATLAB with \textit{Gurobi} as the QP solver (see \cite{Gurobi2014}). The state transition matrices are computed numerically by solving the ODE system \eqref{eq:LTV_transition_matrix} with the \textit{ode45} routine of MATLAB which implements a 4th order Runge-Kutta method with a variable time step. For the continuous thrust case, the second term of the right hand-side of Eq.\eqref{eq:LTV_system_general_solution} is computed with a trapezoidal method integration. Although numerical integrations augments the computational burden, especially the one concerning the transition matrix, in practice these matrices can be computed on ground and uplinked to the probe before starting the manoeuvre. The common conditions for both scenarios are shown in Table \ref{common_scenario_parameters}. 
\begin{table}[h]  
	\centering
	\begin{tabular}{|c|c|c|c|}
		\hline
		$t_0$ & 1d 7h 9m 10s & $t_f$ & 1d 19h 9m 10s \\ \hline
		$c_y$ & 1/$\tan(\pi/6)$ & $c_z$ & 1/$\tan(\pi/6)$ \\ \hline
		$y_0$ & 5 m & $z_0$ & 5 m \\ \hline 
		$\pmb{\delta} \pmb{\bar{\theta}}$ & [2.5$^{\circ}$, 2.5$^{\circ}$, 2.5$^{\circ}$]$^T$ & $\pmb{\Sigma}_{\delta \theta}$ & (2.5$^{\circ}$)$^2$$\mathbf{I}$\\ \hline
	\end{tabular}
	\caption{Global simulation conditions}
	\label{common_scenario_parameters}
\end{table}
Note that a docking sensor has a cone half-angle of $30^{\circ}$. The controller tuning parameters are taken, for both scenarios, as $N=40$, $\gamma=10^6$, $\alpha=0.95$ and $\lambda=0.25$. On the other hand, the specific continuous thrust parameters are chosen as $n_u=400$, $q=4$ and $n_c=44$, hence assuring $C^4$ continuity. Since the disturbances evolve stochastically, 100 random realizations of them, see Eq.\eqref{eq:perturbations_dV}-\eqref{eq:perturbations_continuous}, are simulated. By doing this, the proposed robust controller can be effectively compared with a non-robust one ($\pmb{\delta}_S=\mathbf{0}$).

\subsubsection{Impulsive scenario}

Consider the impulsive scenario defined by Table \ref{dV_scenario_parameters}.
\begin{table}[h]  
	\centering
	\begin{tabular}{|c|c|}
		\hline
		$\mathbf{r}_0$ & [400, 200, -200]$^T$ m\\\hline
		 $\mathbf{v}_0$ & [0.1, -0.1, 0.1]$^T$ m/s\\ \hline
		$\Delta \mathbf{V}_{\text{max}}$ & [0.1, 0.1, 0.1]$^T$ m/s\\ \hline
		$\mathbf{u}_{\text{max}}$ & [0, 0, 0]$^T$ m/s$^2$\\ \hline 
		$\text{max}(\pmb{\delta}\mathbf{\bar{V}})$ & 5$\cdot$10$^{-4}$$\cdot$[1, 1, 1]$^T$ m/s\\ \hline
		 $\pmb{\Sigma}_{\delta V}$ & (5$\cdot$10$^{-4}$)$^2$$\mathbf{I}$ m$^2$/s$^2$ \\ \hline
	\end{tabular}
	\caption{Impulsive scenario conditions}
	\label{dV_scenario_parameters}
\end{table}
The thrust level bias could potentially influence the results. As a consequence, it is considered to be in the interval $\pmb{\delta}\mathbf{\bar{V}} \in$$[-\text{max}(\pmb{\delta}\mathbf{\bar{V}}), \text{max}(\pmb{\delta}\mathbf{\bar{V}})]$ with constant probability. Note that the continuous thrusters are not operative since their bounds are null. 

\begin{figure}[] 
	\begin{center}
		\includegraphics[width=9cm,height=9cm,keepaspectratio]{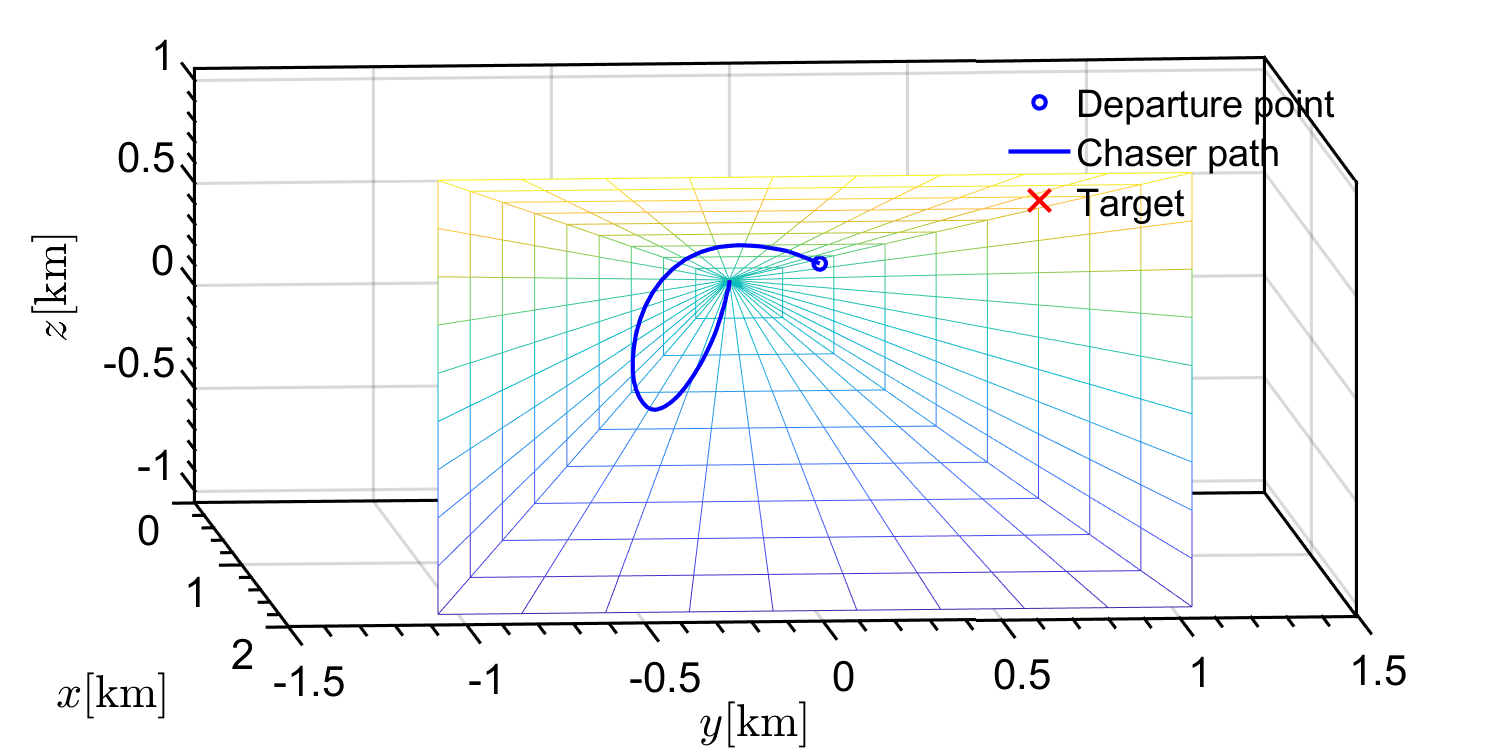}
	\end{center}
	\caption{Chaser 3D trajectory for the first random realization using the robust controller.}
	\label{fig:3D_path_robust_dV}
\end{figure}
\begin{figure}[] 
	\begin{center}
		\includegraphics[width=9.5cm,height=9.5cm,keepaspectratio]{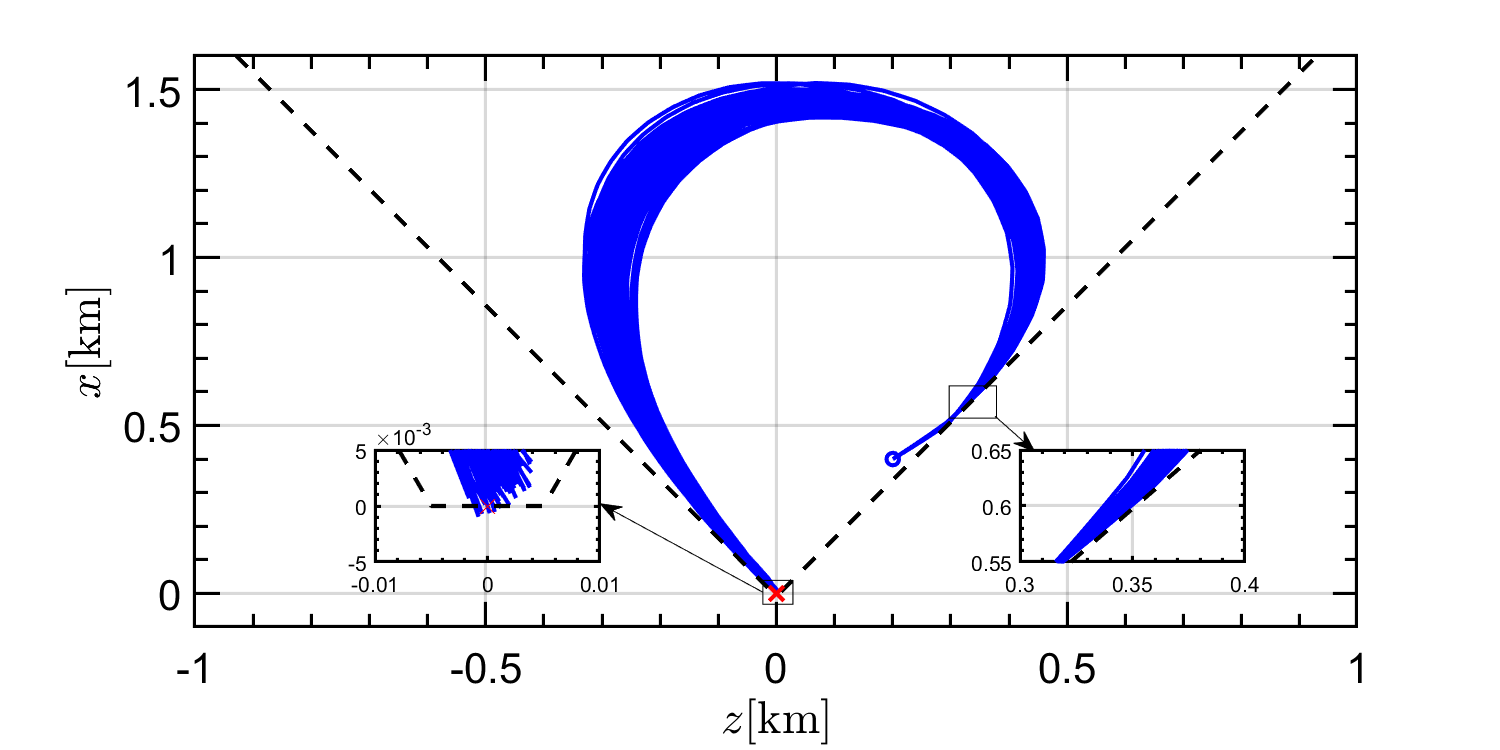}
	\end{center}
	\caption{XZ plane trajectories for all the random realizations using the robust controller.}
	\label{fig:2D_trajectories_robust_dV}
\end{figure}
\begin{figure}[] 
	\begin{center}
		\includegraphics[width=8.5cm,height=8.5cm,keepaspectratio]{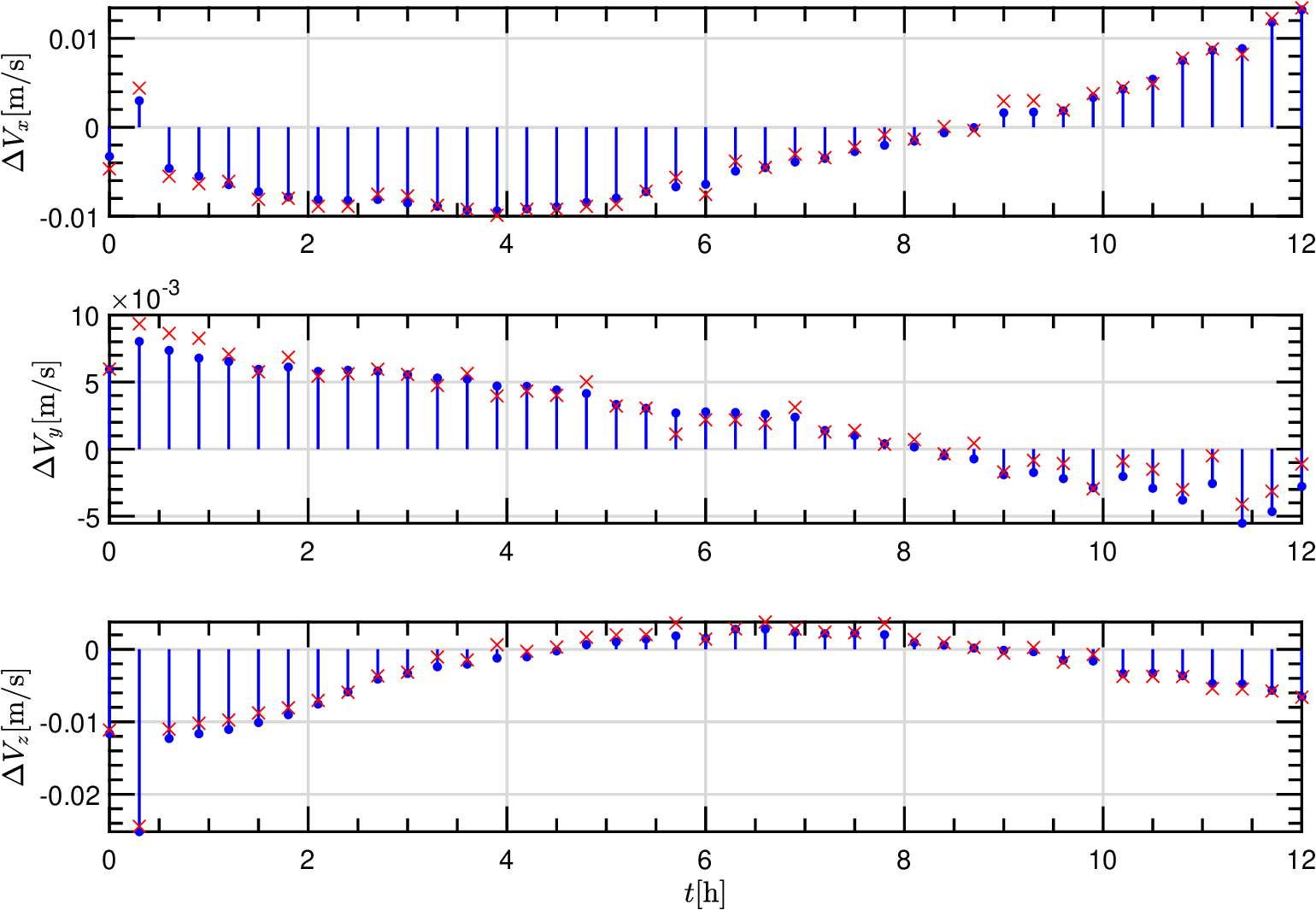}
	\end{center}
	\caption{Impulses for the first random realization using the robust controller. Blue: computed impulses; red: applied impulses.}
	\label{fig:Impulses_robust_dV}
\end{figure}
\begin{figure}[] 
	\begin{center}
		\includegraphics[width=9.5cm,height=9.5cm, keepaspectratio]{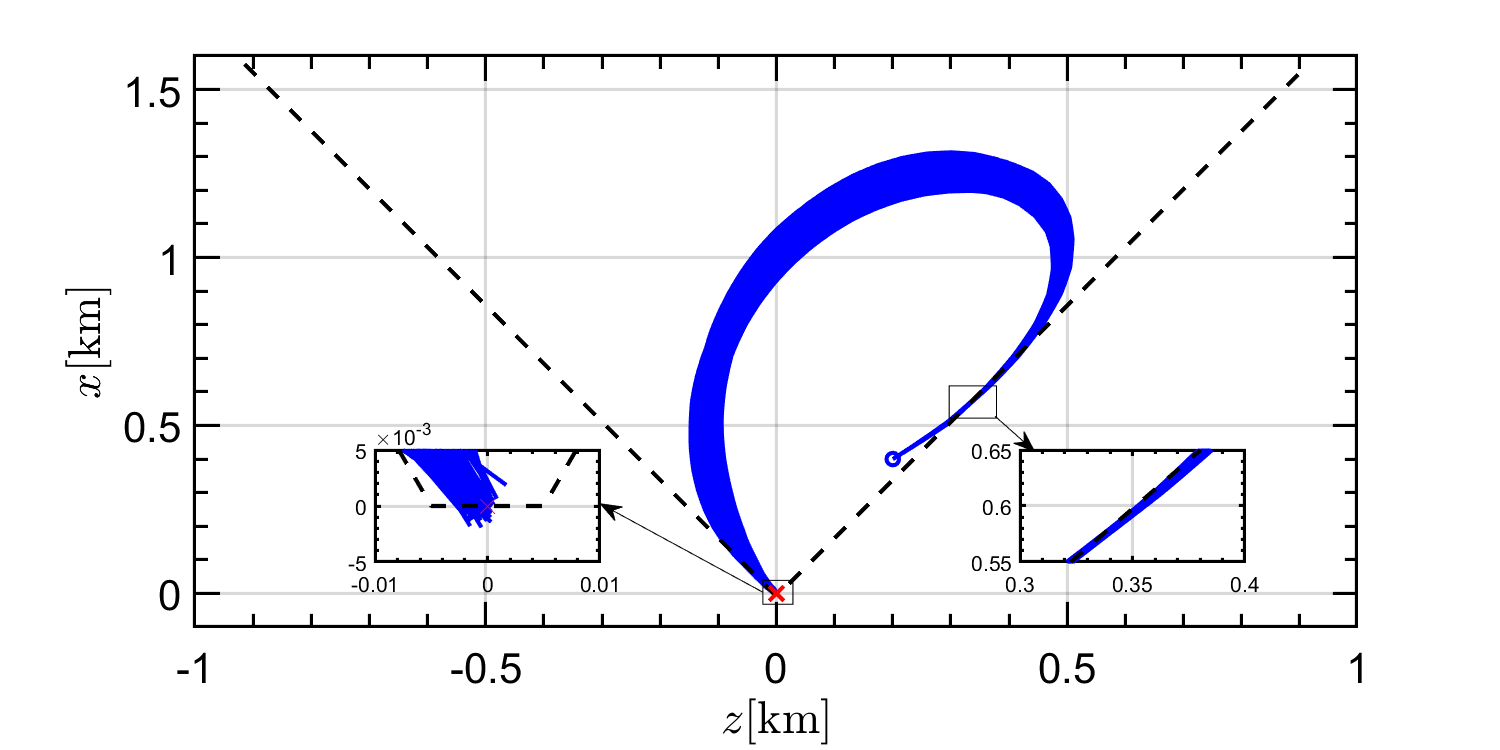}
	\end{center}
	\caption{XZ plane trajectories for all the random realizations using the non-robust controller.}
	\label{fig:2D_trajectories_nominal_dV}
\end{figure}
\begin{figure}[] 
	\begin{center}
		\includegraphics[width=8cm,height=8cm, keepaspectratio]{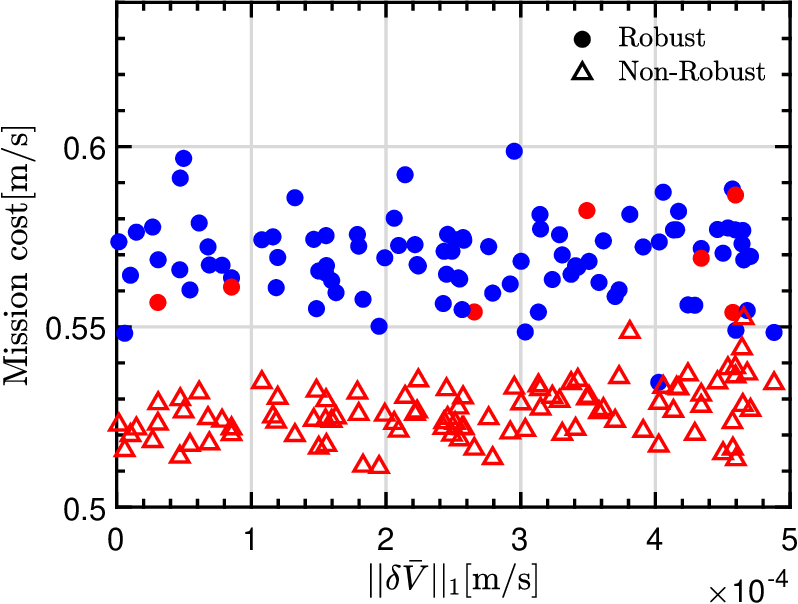}
	\end{center}
	\caption{Mission cost against the thrust level bias for all the random realizations. Blue: LOS satisfaction; red: LOS violation.}
	\label{fig:cost_dV}
\end{figure}

The robust controller simulation results are shown in Fig.\ref{fig:3D_path_robust_dV}-\ref{fig:Impulses_robust_dV}. For the sake of clarity, only the trajectory on the $XZ$ plane is shown, since that projection depicts the most critical LOS constraints. Although, a close range rendezvous scenario is considered, the trajectory start diverging from the target up to $1.5~\text{km}$ approximately, (see Fig.\ref{fig:3D_path_robust_dV}-\ref{fig:2D_trajectories_robust_dV}), which highlights the capability of the linear model to provide fair accuracy at distances above the typical rendezvous ones ($<1~\text{km}$). The final in-track impulses $\Delta V_x$, see Fig.\ref{fig:Impulses_robust_dV}, are positive to brake the chaser and avoid collision with the target. Two critical moments happen along the manoeuvre, the first one taking place just after the departure and the other one at the end of the rendezvous operation, see the zoomed areas of Fig.\ref{fig:2D_trajectories_robust_dV} and Fig.\ref{fig:2D_trajectories_nominal_dV}. Note that the non-robust controller is not capable of guaranteeing LOS constraint satisfaction in any case, see Fig.\ref{fig:2D_trajectories_nominal_dV}, whereas the robust controller avoid the two arising conflicts for the 93\% of the cases, see Fig.\ref{fig:2D_trajectories_robust_dV}. However, in exchange for the safeness increment, the mission cost also increases when comparing the robust approach with the non-robust one, see Fig.\ref{fig:cost_dV}. The computation times, for a i7-860 CPU at $2.80~\text{GHz}$, are of $2.3451~\text{s}$ to compute the stack matrices whereas each MPC step requires $0.4606~\text{s}$ in average requiring the worst case $0.6911~\text{s}$.
  
\subsubsection{Continuous thrust scenario}

Consider the continuous thrust scenario characterized by Table \ref{Continuous_thrust_scenario_parameters}. 
\begin{table}[h]  
	\centering
	\begin{tabular}{|c|c|}
		\hline
		$\mathbf{r}_0$ & [600, 300, -200]$^T$ m \\ \hline
		$\mathbf{v}_0$ & [0.1, -0.1, 0]$^T$ m/s\\ \hline
		$\Delta \mathbf{V}_{\text{max}}$ & [0, 0, 0]$^T$ m/s\\ \hline
		$\mathbf{u}_{\text{max}}$ & [10$^{-4}$, 10$^{-4}$, 10$^{-4}$]$^T$ m/s$^2$\\ \hline 
		$\text{max}(\bar{\mathbf{u}}_E)$ & 5$\cdot$10$^{-7}$[1, 1, 1]$^T$ m/s$^2$\\ \hline $\pmb{\Sigma}_{\delta u_E}$ & (5$\cdot$10$^{-7}$)$^2$$\mathbf{I}$ m$^2$/s$^4$ \\ \hline
	\end{tabular}
	\caption{Continuous thrust scenario conditions}
	\label{Continuous_thrust_scenario_parameters}
\end{table}
Again, the thrust level bias has been considered to vary with constant probability within the interval $\bar{\mathbf{u}}_E\in[-\text{max}(\bar{\mathbf{u}}_E),\text{max}(\bar{\mathbf{u}}_E)]$. 
\begin{figure}[] 
	\begin{center}
		\includegraphics[width=9cm,height=9cm,keepaspectratio]{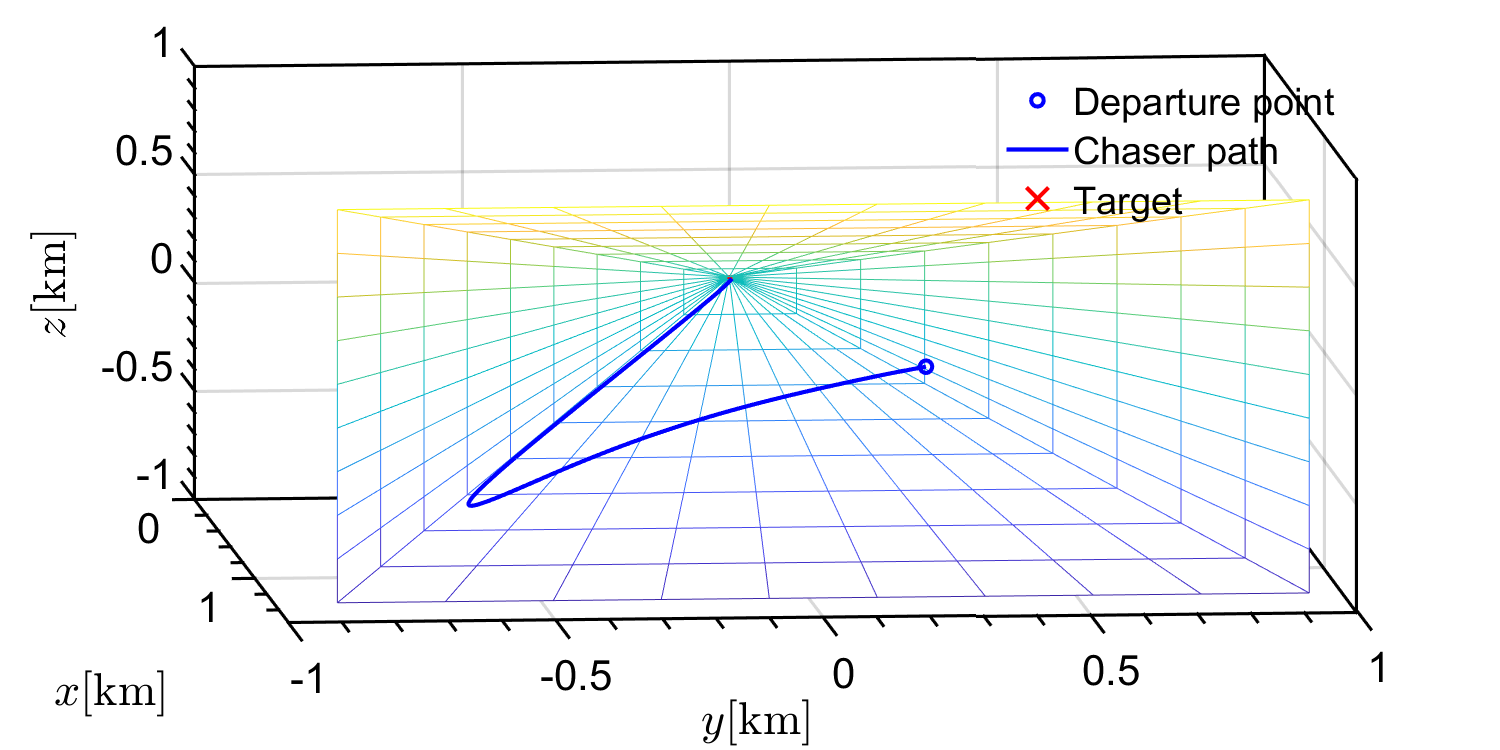}
	\end{center}
	\caption{Chaser 3D trajectory for the first random realization using the robust controller.}
	\label{fig:3D_path_robust_continuous}
\end{figure}
\begin{figure}[] 
	\begin{center}
		\includegraphics[width=9.5cm,height=9.5cm,keepaspectratio]{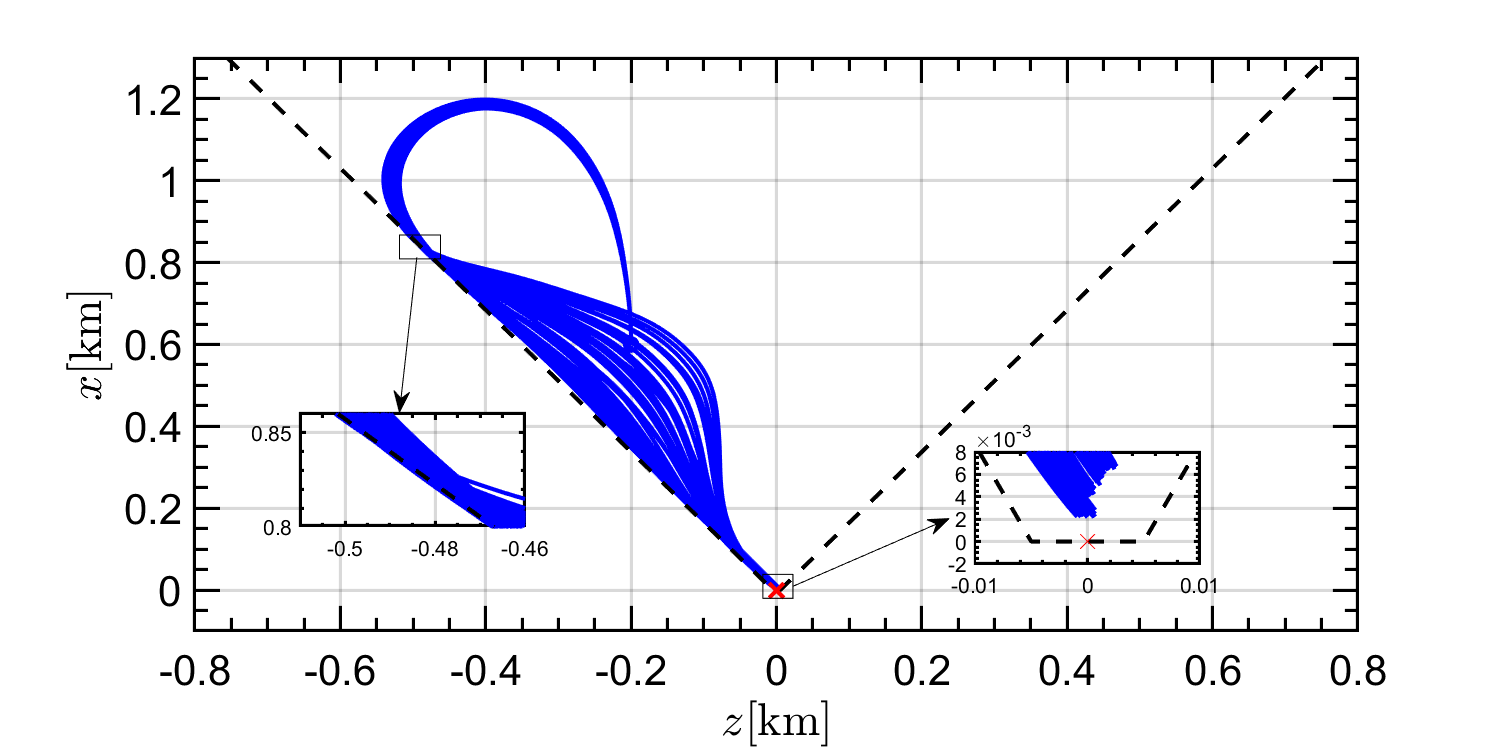}
	\end{center}
	\caption{XZ plane trajectories for all the random realizations using the robust controller.}
	\label{fig:2D_trajectories_robust_continuous}
\end{figure}
\begin{figure}[] 
	\begin{center}
		\includegraphics[width=9.5cm,height=9.5cm,keepaspectratio]{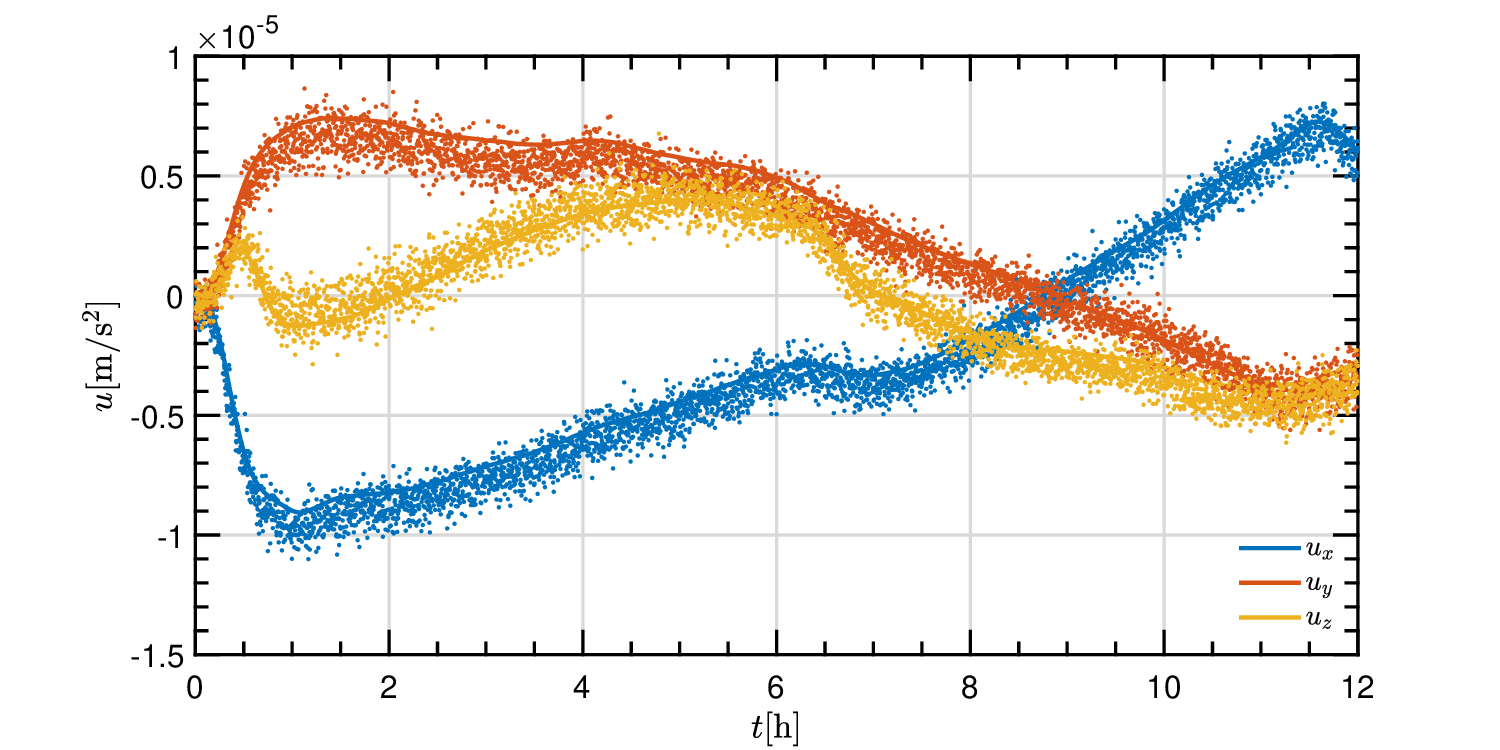}
	\end{center}
	\caption{Thrust acceleration for the first random realization using the robust controller. Solid: computed thrust; dotted: applied thrust.}
	\label{fig:thrust_robust_continuous}
\end{figure}
\begin{figure}[] 
	\begin{center}
		\includegraphics[width=9cm,height=9cm, keepaspectratio]{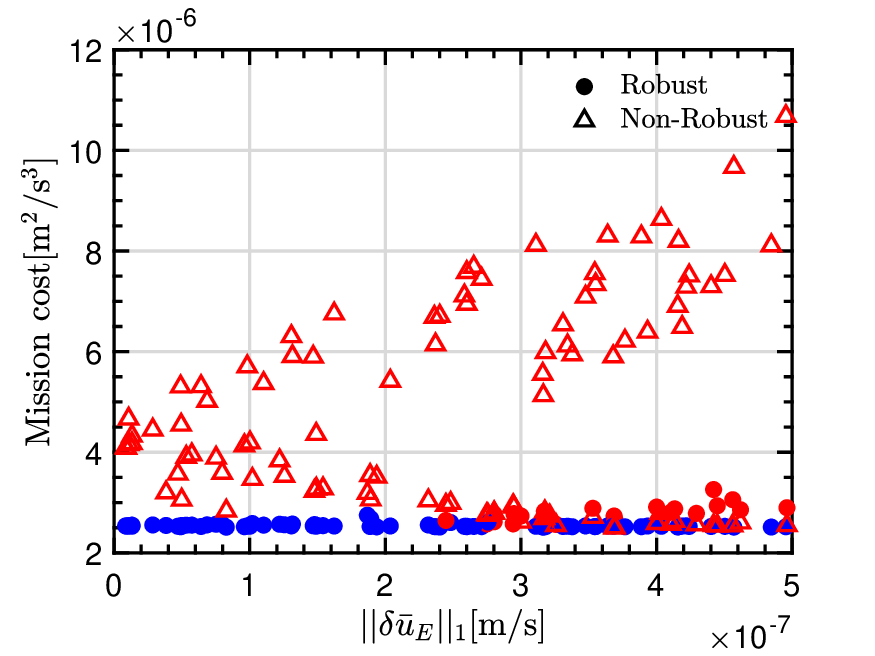}
	\end{center}
	\caption{Mission cost against the thrust level bias for all the random realizations. Blue: LOS satisfaction; red: LOS violation.}
	\label{fig:cost_continuous}
\end{figure}

The robust controller simulation results are shown in Fig.\ref{fig:3D_path_robust_continuous}-\ref{fig:thrust_robust_continuous}. Again, the final control in the in-track direction, $u_x$, is positive to avoid collision with the target, see Fig.\ref{fig:thrust_robust_continuous}. Comparing the robust against the non-robust controller yields again the conclusion that the non-robust one shows worst performance in terms of constraints satisfaction when compared to the chance-constrained method, see Fig.\ref{fig:2D_trajectories_robust_continuous}. As a matter of fact LOS constraint satisfaction is of 76\% (appearing most of the violations at high bias levels, see Fig.\ref{fig:cost_continuous}) for the robust controller whereas the non-robust controller achieves a LOS constraint satisfaction of 1\%. Moreover, in this case, the chance-constrained method consumes less, in general, than the non-robust method, see Fig.\ref{fig:cost_continuous}. Regarding computation times, $6.3504~\text{s}$ are required to compute the stack matrices at the beginning whereas in average each robust MPC step takes $2.0369~\text{s}$ with the most severe computation requiring $3.5514~\text{s}$.    

\section{Conclusions}

In this work, a chance-constrained MPC with disturbance estimation, for restricted three body problem rendezvous, is presented. Moreover, this robust controller is formulated to consider both chemical and electric thrusters, thus increasing the flexibility of the method. The chemical thrusters are modelled as impulsives and the electric ones are parameterized in terms of B-splines. The controller is limited to close rendezvous operations where the system dynamics can be linearized. The simulations have shown a great increase of mission success, sometimes at the expense of the cost, for the robust controller when compared to the non-robust one.  

The main drawback of the algorithm is the numerical integration of the state transition matrices since the dynamics is LTV. However these matrices can be computed by the ground control segment and loaded via uplink to the spacecraft. It is left as future work to evaluate the performance of this algorithm against other robust techniques such as worst-case methodologies, see \cite{Louembet2015}, and tube-based MPC, see \cite{Mammarella2018}. In conclusion, the presented chance-constrained model predictive controller describes an implementable, flexible and relatively fuel efficient algorithm for spacecraft close rendezvous operations in a complex dynamical system under the presence of disturbances.  

\section*{Acknowledgements}
The authors thank Jos\'e Manuel Montilla and Jorge Gal\'an-Vioque for discussions and help with NRHOs.
The authors also gratefully acknowledge financial support from Universidad de Sevilla, through its V-PPI US, and from the Spanish Ministerio de Ciencia, Innovaci\'on y Universidades under grant PGC2018-100680-B-C21.   
 
\bibliographystyle{elsarticle-harv} 
\bibliography{rendezvous_bib}

\end{document}